# The Flexibility and Musculature of the Ostrich Neck: Implications for the Feeding Ecology and Reconstruction of the Sauropoda (Dinosauria: Saurischia)

Matthew J. Cobley

University of Bristol

# The Flexibility and Musculature of the Ostrich Neck: Implications for the Feeding Ecology and Reconstruction of the Sauropoda (Dinosauria: Saurischia)

M. J. Cobley[1]

1 Department of Earth Sciences, University of Bristol, Bristol, UK

Correspondence:

Matthew J. Cobley, Department of Earth Sciences, University of Bristol, Bristol BS8 1RJ, UK.

Email: cobley.mj@gmail.com

Page heading title: Ostrich neck flexibility and musculature: Implications for sauropods

**Abstract**

The Sauropoda were the largest terrestrial animals ever to have lived on this planet. As their nutritional requirements were so huge, their diet holds sway over the ecology of many Mesozoic herbivores. The diet of the sauropods is limited by their feeding envelope, which in turn is governed by the posture and flexibility of their elongate necks. Yet the exact nature of the flexibility and posture of the neck has been a contentious issue. Previous studies have utilised computer models of dry bone, mechanical principles or the flexibility of the necks of extant animals. However, the effect of the musculature of the neck has yet to be investigated. Through measurements of the flexibility of the ostrich neck after cumulative tissue removal, analyses of the muscle attachment sites of the ostrich and sauropods, and testing of the Osteological Neutral Pose model, this study attempts to rectify this situation. The ostrich neck was shown to have three sections of flexibility; a slightly flexible anterior section, a very flexible middle section and a stiff posterior section. The Osteological Neutral Pose did not show these sections, and was shown to potentially overestimate and underestimate flexibility. It was also found that the inter-vertebral space could account for varying estimates of flexibility, and that sauropods would have proportionally more muscle mass at the base of the neck in relation to the ostrich. Ultimately, it was shown that the tissues of the neck place the limits of flexibility, and that zygapophyseal overlap does not indicate the flexibility of the neck. Should the Osteological Neutral Pose affect sauropod flexibility estimates in the same manner as that of the ostrich (a general overestimate), then the sauropods would have a more limited feeding envelope than previously thought, allowing for greater niche partitioning between groups.

**Acknowledgements**

Initial thanks go to my supervisors Emily Rayfield and Paul Barrett for their patience, support and contributions throughout the completion of this project. I am forever grateful for the opportunity to contribute to a subject that has fascinated me for so long.

Special recognition also goes to Remmert Schouten and Mike Taylor. I doubt this thesis would have been completed without their advice and encouragement. I would also like to thank the sauropod-neck community as a whole: Matt Wedel, Darren Naish, Gordon Dzemski and Kent Stevens have all assisted my work and kindly answered any and all questions I have asked.

For their help in the laboratory, appreciation goes to Pedro Viegas, Suzanne Cobley, Sarah Stephens and Anthony Hancy. For their advice and helpful comments on the write-up, I would like to thank Kate Davis.

Additionally I would like to thank my friends and family for supporting me with my work, especially Steven and Mary Cobley, Melissa Johnson, Robert Bick, Daniel Finn and Leigh Maddocks, who all made the process much easier for me.

Final thanks go to Michael Kendrick and William Davies, without whom none of this would have been possible.

**Declaration**

I declare that the work in this dissertation was carried out in accordance with the requirements of the University's Regulations and Code of Practice for Taught Postgraduate Programmes and that it has not been submitted for any other academic award. Except where indicated by specific reference in the text, this work is my own work. Work done in collaboration with, or with the assistance of others, is indicated as such. I have identified all material in this dissertation which is not my own work through appropriate referencing and acknowledgement. Where I have quoted from the work of others, I have included the source in the references/ bibliography. Any views expressed in the dissertation are those of the author.

Signed ……………………………………

Date ……………………………………

**Table of Contents**

## Introduction

The sauropods are unequivocally the largest terrestrial animals ever to have existed. A group of saurischian dinosaurs, the clade Sauropoda was immensely successful from the Late Triassic to the very end of the Cretaceous, with representatives found on all continents (Sander *et al*, 2010). Whilst their general morphology is well understood, the issue of neck posture is still contentious. Some recent studies have proposed the long necks of sauropods evolved for sexual selection (Senter, 2006), however the lack of evidence for this theory (Taylor *et al*, 2011) reinforces the long held view has been that they evolved for maximising the feeding envelope; either for high browsing (Bakker, 1971; Paul, 1987; Christian, 2010) or a wider lateral range of low browsing (Martin, 1987; Ruxton & Wilkinson, 2011). Various theories on the posture and flexibility of the neck have been presented (Stevens & Parrish, 1999; Dzemski & Christian, 2007; Taylor, Wedel & Naish, 2009), with differing approaches leading to various implications for overall biology and ecology. Whilst heart size and output (Seymour, 2009a; 2009b), the structure of the respiratory system (Perry *et al*, 2009; Perry, Breuer & Pajor, 2011), risk of predation and intra-species niche partitioning (Stevens & Parrish, 2005a) are all affected by the position of the cervical column, there are also massive implications for the diet and ecology of the sauropods, and therefore the ecology of many other creatures that co-existed with them during the Mesozoic. Whilst neck posture and flexibility in most species has relatively little effect on their ecology, due to them having relatively short necks, sauropod necks can reach as long as 14 m (Wedel & Cifelli, 2005), meaning smaller differences in the angle the neck is held at lead to differing head heights of a metre or more. The Sauropoda display a wide array of morphologies, but broadly speaking if they were to have roughly horizontal, downward sloping necks, their heads would reach ~4 m high (Stevens & Parrish, 2005b), whilst a more vertical posture would lead to some species with head heights of 16-20 m (Holtz & Rey, 2007). Establishing the flexibility of a sauropod neck allows us to estimate the ‘feeding envelope’ of a given species. This envelope is the maximum range over which an individual could feed, and along with previous work on the flora present at the time (Chin, 1997; Hummel *et al*, 2008; Gee, 2011) and sauropod dentition (Calvo, 1994; Fiorillo, 1998; Upchurch & Barrett, 2004; Stevens & Parrish, 2005a), allows us to identify the diet of the sauropods. Establishing their diet is extremely important; as the sauropods were so large, requiring anywhere up to 400kg of dry plant matter per day for an adult (Hummel *et al*, 2008), it is not hard to imagine a herd of these animals stripping an area of vegetation in a short amount of time. Reducing the resources in a given area would force other species present to adapt by either feeding on different material, or through temporal or spatial niche partitioning of the same plants.

Previous work on sauropod neck posture and flexibility has led to three general theories. The first method, through computer modelling of the neck, was based on the assumption that the vertebrae of the neck could not be flexed past the point where there was a minimum of 50% overlap of the zygapophyses of two adjacent vertebrae: The ‘Osteological Neutral Pose’ (ONP) (Stevens & Parrish, 1999). This leads to estimates of low flexibility in sauropod necks, and the conclusion that species such as *Diplodocus* and *Apatosaurus* held their necks in a downward sloping fashion, much different from the classical, vertically held depictions (Stevens & Parrish, 1999; Stevens, 2002; Stevens & Parrish, 2005a; 2005b). This work was questioned by studies using direct comparisons with the posture held by extant species, asserting that all extant amniotes held their heads in a vertical fashion, and as such it was most parsimonious to reconstruct sauropods with a swan-like, ‘S’- shaped posture (Taylor *et al*, 2009). Mechanical models have also been implemented, which leads to a middle ground between these two theories; with the neck being held slightly above horizontal with a reasonable amount of flexibility (Preuschoft, 1976; Alexander, 1985; Christian & Preuschoft, 1996; Christian & Heinrich, 1998; Christian, 2002; Christian & Dzemski, 2007; 2011). Studies quantifying the flexibility of extant necks also come to this conclusion; Dzemski & Christian (2007) studied the flexibility of *Struthio camelus* (the ostrich), *Giraffa camelopardis* (the giraffe) and *Camelus bactrianus* (the Bactrian camel), with all tissues intact and solely the neck skeleton.

However, none of these previous studies have analysed the effects of tissue on the flexibility of the skeleton; the ONP relies solely on bone to make its estimates (Stevens & Parrish, 1999); Taylor *et al* (2009) use the neck as a whole as a more superficial means of comparison; the 'Preuschoft method' (Christian & Dzemski, 2007; 2011) deals solely in the mechanics of the neck. Studies based on the flexibility of extant animal necks have yet to study the actual effect of tissues on the flexibility of the neck, instead comparing the flexibility of the whole neck and that of the neck skeleton. This study aims to rectify this situation. By measuring the flexibility of the neck with sequential and cumulative removal of tissues, a picture of how tissues of different sizes and placements around the neck affect flexibility will become apparent. By measuring the attachment site of various muscles in the ostrich neck and in sauropods, we can also attempt to estimate the relative amounts of muscle mass around the necks of the extinct species (Sniveley & Russell, 2007a). Where previous studies have mainly focused on the ONP as a predictor of posture (Christian & Dzemski, 2007; Taylor *et al*, 2009), this study will analyse the theory's potential for estimating maximum flexibility of the neck. The effect of cartilage will also be investigated; whilst the muscles of the neck can be inferred, their mass and placement within the neck are debateable. The presence of cartilage is much less contentious, yet is something that previous studies have not accounted for. The study will be conducted using the ostrich as a representative from the 'extant phylogenetic bracket' (EPB) (Bryant & Russell, 1992; Witmer, 1995), and as it is the most commonly used avian in previous studies (Christian & Dzemski, 2007; Dzemski & Christian, 2007). These analyses will be brought together to assess the feeding envelope of sauropods, and critique previous methods of estimating posture and flexibility.

Institutional abbreviations: NSMT: National Science Museum, Tokyo; CM: Carnegie Museum of Natural History, Pittsburgh. Other abbreviations: EPB: Extant Phylogenetic Bracket; ONP: Osteological Neutral Pose. Abbreviations of muscle attachment sites listed in Table 1.

## Methodology

### Animals studied

*S. camelus* has been chosen as an analogue for the sauropod neck using the EPB approach (Bryant & Russell, 1992; Witmer, 1995). As the Sauropoda are stem Avians, and the Struthioniformes are the largest birds to exhibit elongate necks, ostriches are a suitable candidate for comparative study. Thought the ratites have evolved elongate necks independently several times (van der Leeuw, Bout & Zweers. 2001), they are a more viable candidate for study than mammals, due to their bracketing of the sauropods, and their relatively more similar number of cervical vertebrae; mammals are limited to seven (Galis, 1999). Three female ostrich necks were used in this study, obtained from MNS Ostriches Ltd, U.K. All three were humanely destroyed at around the same age (~ 6 months). All three necks had been separated from the torso prior to being obtained; two had been pre-skinned and decapitated, whilst one had its head and skin intact. The necks were frozen immediately after amputation to minimize decomposition, and frozen for a sufficient amount of time so that rigor mortis would no longer have an effect.

Analysis of flexibility of the cervical column

The necks were examined immediately after thawing. Detailed notes, sketches and digital photographs were made of the muscles and tissues present in the neck of the ostrich. The flexibility of the neck was measured at various stages of cumulative tissue removal: (In sequential order) with all tissue intact, after removal of the long dorsal musculature (M. biventer cervicis; M. longus colli dorsalis; M. ascendens cervicalis), after removal of the long ventral musculature (M. flexor colli medialis; M. longus colli ventralis); after removal of the lateral musculature (M. flexor colli lateralis) after removal of the single-segment muscles (Mm. intercristales; Mm. interspinales; Mm. intertransversarii); after removal of the ligamentum elasticum. These groups are based on the placement of the muscle in relation to the vertebrae rather than their function. Flexibility measurements were made using a medical goniometer, measuring the flexibility about each inter-vertebral joint, where flexibility amounted to the degree of movement a given vertebra was capable of in relation to the vertebra immediately posterior (Fig. 1). All flexibility measurements are given as deviations from 0°, where the anterior vertebra is angled in a straight line with the posterior vertebra. Should the vertebra not align at 0° i.e. if they are unable through natural dorsiflexion, there would be little impact on the measurements as 0° is essentially in line with the posterior vertebra. The mass of removed muscle and other tissues, in addition to the mass of the neck, was measured with a high precision scale after each stage of removal ((0–810g) Ohaus, d=0.01g; (>810g) Ultraship, d=1g). One neck was separated into 15 sections at each intervertebral joint. The mass of each section was measured, and all tissues were removed. After this the mass of the vertebra was measured to give the mass of tissue around each vertebra. The mass of tissue surrounding each inter-vertebral joint was then estimated using half the mass from the vertebra anterior and half from the vertebra posterior to a given joint. Each neck was cleaned of all soft tissue by being boiled several times in water until all tissue and fat was removed. Measurements of neck length along the most dorsal edge of the neck were taken before and after removal of the tissue with a tape measure. The centra were measured with a tape measure immediately after the boiling process whilst the cartilage was still wet; after being left to dry; and after removal of the cartilage with a scalpel

Proportional variation of muscle attachment sites

The size of the muscle attachment sites on the vertebrae of the ostrich was measured using digital photographs and the freeware computer programme ImageJ (Abramoff, Magalhaes & Ram, 2004). The attachment sites measured were; the Ansa costoransversaria; crista transverso-obliqua; cristae laterales; processus caroticus; processus costalis; processus spinosus; tubercula ansae; torus dorsalis (Fig. 2). The muscles originating from these attachment sites were identified during the dissections. Digital photographs of two sauropods necks were also measured using ImageJ; one of the fossil remains of *Apatosaurus ajax* (NSMT-PV 20375) (Upchurch, Tomida & Barrett, 2004), and casts of *Diplodocus carnegii* (CM-84). Photographs of *D. carnegii* were taken of casts at the Museum fur Naturkunde

Zentralinstitut der Humboldt-Universitat zu Berlin, Germany, and are available online (Dzemski, 2005). In addition to the above attachment sites, the size of the spinopostzygapophyseal lamina was also measured (equivalent to the crista transverso-obliqua (Wedel & Sanders, 2002)). These sites were then converted into proportions relative to the length of the neck and of the respective centra, allowing us to see how the size of a given attachment site changes across the whole neck. This is similar to a previous study by Sniveley & Russell (2007a), which used the origin sites of muscles to compare the cross-sectional area of muscles in theropods. Sniveley & Russell (2007a) used the length of the entire neck as its scale of proportion; however as this analysis concerns changing flexibility along the neck, rather than flexibility as a whole, the proportion of the attachment sites in relation to length of the associated centrum will also be analysed. Centrum length was measured using ImageJ as with the attachment sites. The length of the neck was measured directly from the ostrich specimen. For the sauropods the estimated length for the *Apatosaurus* specimen given by Upchurch *et al* (2004) was used, and an estimate was taken from scale drawings of *Diplodocus* specimen CM-84 (Hatcher, 1901). Though not giving exact figures as to the amount of muscle originating from each site, or the flexibility allowed, the data will show the relative difference in muscle mass in sauropods compared with the ostrich.

## Osteological Neutral Pose

A series of analyses were completed to test the hypothesis that the flexibility of extant animal necks could be predicted by the ONP (Stevens & Parrish, 1999). Allowing a minimum of 50% overlap for dorsoventral and lateral movement, and using 100% overlap as a ‘resting’ position, the maximum degree of flexibility was measured for the ostrich neck skeleton whilst the cartilage was wet (immediately after boiling off the soft tissue); after drying the cartilage; and after removal of the cartilage. By taking the degree of flexibility at 50% zygapophyseal overlap dorsally and ventrally, we can calculate the degree of flexion allowed per 1% change in overlap (1). Applying this to the maximum flexibility values measured from the neck with all tissues intact, we can estimate the actual overlap exhibited during flexion of the complete neck (2).

(Dorsal flexibility at 50% overlap + Ventral Flexibility at 50%) / 100 = Degrees of flexion per 1% change in zygapophyseal overlap (1)

Actual Flexibility / ° per 1% = Actual Overlap (2)

Naming conventions used

Due to the complex nature of the cervical musculature and a previous lack of consensus over the naming of the various muscles, it is important to state the conventions used for the naming of the various muscles and muscle attachment sites. Recently the terms for musculature of avians have begun to stabilize after the wider implementation of the Nomina Anatomica Avium (Baumel *et al*, 1993); as such this will be used as the basis for the naming of the avian musculature. As it is the only paper to explicitly explore the homologous muscle attachment sites (and musculature) of extant avians and sauropods, the naming of the various attachment sites will follow Wilson (1999), congruent with previous studies concerning homologous attachment sites (Wedel & Sanders, 2002), however a full description of the location of these attachment sites is provided when necessary.

## Results

Systematic reconstruction of the tissues present in the ostrich neck

The muscles of the neck and their respective attachment sites were observed in the necks of the ostrich (Table 2).

M. biventer cervicis (m. biv. cerv.)

Origin: Neural spines of the posterior-most cervical vertebrae, or anterior-most caudal vertebrae

Insertion: Parietals

Function: Though the muscle does not 'attach' to any point of the necks studied (the ostriches were both decapitated and separated from the body at the base of the neck), the two bellies of m. biv. cerv. are nonetheless present. These bellies are present within the same sheath of fascia as m. long. col. dors.. The bellies taper gradually to C8, connecting to a pair of tendons that are the dorsal-most tissues of much of the neck (barring connective tissue and skin) (Fig. 3). These tendons run to the base of the head where another paired set of muscular bellies are present. These were not observed in the specimens studied due to aforementioned decapitations.

M. longus colli dorsalis (m. long. col. dors.)

Origin: Processus spinosus – Aponeurosis notarii, from neural arches and transverse processes of the posterior-most cervical vertebrae (Fig. 4).

Insertion: Torus dorsalis – Slips insert on the dorsal processes alongside m. ascendens cervicalis.

Function: Like the m. biv. cerv., m. long. col. dors. consists of a large amount of muscle mass at the base of the neck, connected to the anterior portion of the muscle complex (present around C1-C3) by tendons. The muscles are bound in the same fascial sheath as m. biv. cerv.. The muscle is exclusively used for dorsiflexion of the neck, especially raising of the anterior vertebrae relative to the base of the neck.

M. ascendens cervicalis (m. asc. cerv.)
Origin: Ansa costotransversaria
Insertion: Torus dorsalis
Function: The m. asc. cerv. runs from the ansa costotransversaria of the posterior vertebra of the posterior vertebra to the torus dorsalis of the second anterior-most vertebra relative to its origin (Fig. 4). Though positioned lateral to the bypassed vertebra (and therefore the centre of rotation), the dorsal position of the anterior insertion allows this muscle to act during dorsiflexion.

M. flexor colli lateralis (m. flex. col. lat.)
Origin: Tubercula ansae, cristae laterales
Insertion: Processus costalis
Function: As with m. asc. cerv. the insertion of this muscle is on the lateral parts of the vertebrae (here the lateral tubercules rather than the ansa costotransversaria). While m. asc. cerv. runs lateral and dorsal to the centre of rotation, m. flex. col. lat. inserts ventrally at the cervical rib. This muscle is used primarily for ventriflexion, however due to the lack of long lateral muscles in the avian neck it is likely that it also aids in lateral flexion when simultaneously flexing downwards.

M. flexor colli medialis (m. flex. col. med.)

Origin: Processus caroticus, processus costalis

Insertion: Processus ventralis corporis, processus costalis

Function: Unlike m. flex. col. lat., m. flex. col. med. runs solely along the ventral portion of the neck. Positioned axial to the cervical ribs, the muscle has no input with regards to lateral flexibility and is only utilised for ventral excursions.

M. longus colli ventralis (m. long. col. ven.)

Origin: Processus caroticus, processus ventralis corporis

Insertion: Processus costalis

Function: Multiple slips of m. long. col. ven. can arise from the same attachment site. This allows for complex ventriflexion (Sniveley & Russell, 2007b). There is a reduction in the number of slips arising from the vertebrae closer to the head compared to vertebrae at the base of the neck. In addition to ventriflexion, the muscles also prevent damage to the neck during dorsiflexion by acting as a damper (van der Leeuw et al, 2001).

Mm. intercristales

Origin: Crista transverso-obliqua

Insertion: Crista transverso-obliqua of the immediately anterior vertebra

Function: These muscles run from the dorsal surface of one vertebra to the adjoining anterior vertebra. This allows for intervertebral dorsiflexion of individual intervertebral joints. Towards the base of the neck these muscles make up far less of the total muscle mass than they do at the anterior, and due to the increased moment arm that is being raised, it is likely that the mm. intercristales also take up a function in stabilising these joints rather than flexing the neck.

Mm. interspinales

Origin: Processus spinosus

Insertion: Processus spinosus

Function: Like mm. intercristales, these single-segment muscles run dorsally along the neck between the neural spines of adjacent vertebrae. These are less well defined than the mm. intercristales, and act to stabilise the joints of the neck as they are too small to have a major impact in dorsal flexion.

Mm. intertransversarii (mm. intertrans.)

Origin: Tubercula ansae, cristae laterales

Insertion: Tubercula ansae, cristae laterales

Function: The muscles both originate and insert at the same attachment sites on the lateral tubercles of adjacent vertebrae (Fig. 4). There is disparity in published literature as to whether the origin or the muscle is on the anterior or posterior of any two vertebrae (Sniveley & Russell, 2007b). However as the smaller of the two vertebrae is most likely to be affected by any contraction or relaxation of mm. intertrans., this study will treat the posterior vertebrae as the origin. Due to the lateral placement of the muscle, contraction of mm. intertrans. leads to lateral flexion to either side of the neck. Though these are short-segmented, inter-vertebral muscles, they are the most important for lateral flexion as there are no laterally flexing long muscles (spanning three or more vertebra) present in avian necks.

Ligamentum elasticum

The ligamentum elasticum is a series of short ligaments present between all vertebrae, adjoining the dorsal processes. The ligament prevents extreme ventral excursions in adjacent vertebrae, and its presence may also prevent dorsal excursions.

Ligamentum nuchae

The ligamentum nuchae is an elastic sheath that surrounds much of the liamentum elasticum. It prevents extreme ventral excursions across the whole neck.

Skin

The skin surrounding the ostrich neck is extremely loose, allowing for the large dorsal and ventral excursions seen in live animals. This prevented any accurate measurement of flexibility between individual vertebral pairs, as the degree of flexibility between the two vertebrae was not conveyed by the skin. This was not the case when the entire neck was dorsally or ventrally flexed, however this study is primarily concerned with the flexibility between individual intervertebral joints.

Flexibility

The maximum dorsoventral flexibility of the ostrich neck after sequential and cumulative removal of muscles was measured (Fig. 5). The flexibility of the ostrich neck with all muscles intact can be divided into 3 sections (Fig.5a). Between C3 – C6, with dorsal extension reaching 12°-19°, C7-C11, with dorsal extension peaking at 25.6° and ranging down to 19.6°, and the posterior section C12-C15, with dorsal extension ranging from 13-15°. Ventral flexion of the neck does not exhibit the same range as dorsal extension, the maximum excursion from 0° being joint 7 at 15.6°, however the same, three sectioned pattern can be observed, especially in C12-C15, where the vertebrae are unable to flex past 0° and thus in a permanent state of dorsal extension. There is a noticeably larger variation in the ventral flexibilities of the neck in comparison to maximum dorsal excursions. Lateral flexibility follows a similar pattern, with comparatively low values at the anterior end of the neck, increasing to >10° for C5-C10, and then decreasing gradually from C11 to the base of the neck, where there is little flexion (<5°) (Fig. 6b).

Removing the long dorsal muscles of the neck increases flexibility across the whole neck, allowing up to 10° more dorsal flexibility, and up to an extra 6.5° of ventral flexibility (Fig. 5b). With the removal of these muscles the posterior vertebrae become flexible enough to flex ventrally past the midline, aside from C12 which is still limited to 1° of dorsal extension. The three sections of the neck are less apparent when looking at the figures for dorsal extension, however they are still apparent during ventral flexion, though joint 6 appears to be part of the 'mid-section'. Removing the dorsal muscles of the neck leads to an increase in lateral flexibility across the neck, allowing for large excursions from 0° from C3 – C8, though there is still limited flexibility of a maximum of 6° at the base of the neck (Fig. 6b).

Removing the long ventral muscles of the neck again increases the flexibility (Fig. 5c); however this increase is less pronounced than from removal of the dorsal removal, with the highest increase in flexibility being 4° (C3). The three sections of the neck are still apparent, and all vertebrae in the posterior section are capable of ventral flexion. Increased values for lateral flexibility across the whole neck occur after the removal of the ventral musculature (Fig. 6c).

Removal of the lateral muscles of the neck leads to further increases in flexibility, much larger than the increase after removal of the ventral musculature (Fig. 5d). This is especially apparent in ventral flexion, where previously overall ventral flexibility was much lower than that of dorsal flexibility, removal of the lateral musculature leads to comparatively similar flexibility values. However, the ventral flexion capabilities of the posterior section of the neck are still limited, at most reaching 10.5°.  With regards to lateral flexibility, the large differences between the anterior and posterior joints are less apparent after removal of the lateral muscles, with the range reduced to 10o where previously it was 21° (Fig. 6d).

The three sections of the neck are less distinct after removal of the single-segment muscles of the neck, leading to another small increase in flexibility (Fig. 5e). Whilst the posterior section is still apparent in ventral flexion, there appears to be a rise in flexibility between C3 and C7, which then drops from C8 to C11. Removal of the single-segment muscles brings back the observable difference in lateral flexibility between the anterior and posterior portions of the neck, with joints between C3-C8 all exhibiting flexibilities of >15° (maximum 23° – Joint 5), whereas the posterior joints all fall between 11° and 13° (Fig. 6e).

Removal of the ligamentum elasticum leads to a massive increase in ventral flexibility, especially in joints between C5 and C8 (Fig. 5f). There is no longer any observable pattern in dorsal flexibility, with values ranging anywhere between 19° and 32°. Lateral flexibility of the anterior vertebrae slightly decreases, and the posterior vertebrae show a large decrease aside from joint 15, which increases to 25° (Fig. 6f). This is likely due to the measurements being taken from a solitary neck rather than the two or three for the five other stages of measuring.

## Tissue mass & measurements

The masses of the various muscle groups of the neck were measured (Table. 3). M. biv. cerv. is by far the most massive at 253g (23.7% of the total weight), making up a large proportion of all the long dorsal muscles (40.5%). The long ventral and lateral muscles are similar in mass (16.47% and 17.22% respectively), whilst the single-segment muscles make up just over a quarter of the total muscle mass of the neck. The mass of each vertebrae and its associated tissue was also measured (Fig. 7a), and the mass of tissue that surrounds each intervertebral joint estimated (Fig. 7b). Whilst the mass of each vertebra shows a steady increase from C3-C17 (6.72g – 42.19g), there is a sharp increase in tissue mass from C11–C17 (53.65 – 159.68), where there was previously a steady increase in tissue mass from C3-C10 (23.45 – 49.39). On average the tissue associated with each vertebra weighs around 3 times that of the vertebra itself. The mass of tissue around each vertebra follows this same pattern, with a steady increase in mass up to C10, where after the amount of tissue increases dramatically.

Measurements were taken of total length of the dorsal side of the neck before and after tissue removal. Prior to tissue removal the average total length of the neck was 76+/-4.5cm (n=3). After tissue removal, with all centra touching, this length was reduced to 70.1 +/-3.75cm (n=3). Lengths of the individual centra were also measured after boiling off all tissue; whilst still wet, after drying, and after removal of the cartilage caps on each end (Table 4). Drying leads to an average loss of 0.16+/-0.15 cm in centrum length for each vertebra, whilst removal of the cartilage caps leads to an average loss of 0.21+/-0.2 cm.

## Osteological Neutral Pose

Measurements for the ONP in the ostrich neck show there is a trend towards higher dorsal and lower ventral flexibilities towards the posterior end of the neck in the specimens studied (Fig. 8a). When measuring maximum lateral flexibility in the ONP, there is no clear pattern present and large variation in the maximum flexibility of specimens studied (Fig. 8b).The angles of deflection obtained when the vertebrae are positioned in the 'neutral' position i.e. 100% overlap of the pre- and post-zygapophyses of adjacent vertebrae were measured (Fig, 9). Though there are large variations in the results there is a trend towards a larger neutral angle in the posterior-most vertebrae (15-20°), where in the anterior vertebrae this angle is much lower (3°-8°). This neutral position is illustrated in Figure 10, along with maximum dorsal and ventral flexion with a minimum of 50% overlap of the zygapophyses.

The maximum flexibility allowed by the ONP when the cartilage is in different stages of drying was measured (Fig. 11). Dried cartilage allowed slightly more flexibility than wet cartilage (Fig. 11a; 11b). The flexibility of the neck with the cartilage removed from the vertebrae undergoes a large increase in overall flexibility of the neck in comparison with vertebrae with the cartilage present (Fig. 11c).

The amount of overlap between adjacent vertebrae when all muscle tissues were intact was estimated from the amount of flexibility allowed when the neck skeletons were oriented with a minimum of 50% overlap dorsally and ventrally (Fig. 12). Dorsally there is a large amount of overlap between the anterior joints, lowering to a minimum overlap of about 40%, this then rises consistently to a maximum of 70-80% overlap in joints 11-14, with joint 15 showing about 100% overlap between the pre- and post-zygapophyses. There is less variation in ventral overlap across the whole neck; however it does exhibit the same decrease in overlap in the middle section, before an increase in overlap towards the base of the neck. Overall minimum ventral overlap is much higher than that of dorsal overlap.

Proportional variation in muscle attachment sites

The ratio of size of the muscle attachment sites to the length of the centra (Fig.13) and neck (Fig. 14) was measured in the ostrich, *Apatosaurus* and *Diplodocus*)

Ansa costotransversaria

Whilst in the ostrich the size of the ansa costotransversaria is relatively constant in relation to centrum length, there is a massive increase in relative size in *Apatosaurus*, where the ratio rises from 1.961 in C5 to 7.377 in C10, remaining above 7 through to C14. This increase is also present in *Diplodocus* though not as dramatic, remaining below a ratio of 1:1 (0.667-0.815) through C3-C8, before increasing at C11 and C12 (1.427 and 1.112 respectively), and again at C15-C16 (2.785 and 2.205). This same pattern is observable in relation to neck length; with a steep increase observable in the *Apatosaurus* from C5-C10, a steady increase in the *Diplodocus* vertebrae, and a relatively stable proportion over the ostrich neck.

Processus spinosus

Again the ostrich has the lowest attachment site size relative to centrum length, and it varies little over the course of the neck. The *Apatosaurus* processus spinosus drops in relative size from C6 – C9 (0.330-0.167), before rising again to above 0.5 in C12 and C14.The *Diplodocus* processus spinosus trends towards an increase in size over the neck, however there is a large drop off between C14 and C15 (1.381-0.616). In relation to neck length, the three species have similar proportions from C3-C6. Posterior to this the *Diplodocus* shows a sharp increase in size to C8, before levelling off, with a large drop off in relative size is still present in *Diplodocus* cervices C14 and C15. The *Apatosaurus* proportional size decreases slightly to C7 before an increase to C12, then levelling off. The ostrich processus spinosus stays at a comparatively stable relative size, with a slight increase over the course of the whole neck.

Crista transverso-obliqua / Spinopostzygapophyseal lamina

Whilst the size of the crista transverso-obliqua relative to centum length is similar in all three animals from C3-C9, the ostrich shows a trend towards a lower relative size, which levels off to a near constant ratio between C11 and C16. This is in contrast to the *Apatosaurus* vertebrae which increase greatly from C9-C10 (0.240-0.421). The *Diplodocus* spinopostzygapophyseal lamina decreases in size from C8 to C11/C12, however shows a sharp increase in size in C14 and C15. Where the *Diplodocus* shows a small proportional spinopostzygapophyseal lamina size in relation to neck length in the anterior-most vertebrae, before a sharp increase from C5-C8 and then levelling off, the *Apatosaurus* stays relatively constant, with some undulation, to C9, before a jump up in size at C10, then levelling off. The ostrich shows a similar pattern, with an exponential increase between C3 and C6, decreasing from C7 to C9 before a slight increase before a levelling off to C16.

Torus dorsalis

The relative size of the torus dorsalis in comparison to the centrum is much the same as that of the crista transverso-obliqua, with the anterior vertebrae of all three species much the same, but the sauropods showing a gradual trend towards a larger proportional attachment site posterior from C6, whilst the ostrich shows a decrease from C6-C10 before levelling off. Relative to total neck length, the sauropods both show a slight increase in proportional size over the whole neck (*Apatosaurus* C3= 0.008, C14=0.014; *Diplodocus* C3=0.005, C14=0.160). The ostrich is proportionally similar to the sauropods over the course of the whole neck; however there is a steepening decrease from C3 – C9, and then a steady increase to C16.

Tuberculum ansa

Proportionally the size of the tuberculum ansa in relation to centrum length is similar in the anterior-most vertebrae; however from C5 onwards the *Apatosaurus* vertebrae show a steep increase in size, *Diplodocus* showing a gentle increase, and the ostrich decreasing slightly down to C9 before increasing slightly through to C16. The figures for relative neck length show much the same pattern, with similar proportions between C3 and C5, before a steep increase in the *Apatosaurus*, a less steep increase in the *Diplodocus*. However the ostrich keeps relatively stable until C12 before a trend towards an increase in proportional size of the tuberculum ansa attachment site.

Processus costalis

The relative size of the processus costalis follows the same pattern in relation to neck and centrum length, with a trend towards a larger proportional size in both the *Apatosaurus* and the ostrich, however the relative size of the *Apatosaurus* attachment site is much larger than that of the ostrich across the whole neck.

Crista lateralis

The relative size of the crista lateralis of the ostrich, in relation to both the centrum length and neck length, follows no discernable pattern.

Processus caroticus

The processus caroticus of the ostrich shows a steady increase in relative size compared to both centrum length and neck length.

## Discussion

Flexibility

The general pattern of three sections of the neck with varying flexibility concurs with previous research into the flexibility of avian necks (van der Leeuw *et al*, 2001 (Pg. 248, Fig. 2)), where the pattern was observed in smaller birds with elongate necks (*Rhea americana* (rhea) and *Cygnus olor* (Mute swan)), and in birds that did not have relatively long necks. The pattern of flexibility with all tissue intact also mirrors that of previous work on the neck flexibility of ostriches (Dzemski & Christian, 2007 (Pg. 707, Fig. 7a), however maximum flexibility in said study was judged to be much larger than in the research detailed here, with both dorsal and ventral flexibility reaching up to 30° (as opposed to a maximum of 25° dorsal, 15° ventral). The posterior-most vertebrae of the specimens in this study were also incapable of any ventral excursions past the midline of 0°, which is not the case in previous work. However as the same pattern of flexibility is apparent throughout the length of the neck, it is likely the difference is due to the specimens themselves rather than the sampling method. Whilst this study used sub-adult ostriches, adults were used in the previous research. It is possible that the smaller neck of the sub-adult is restricted in its movement, to allow time for the musculature of the neck to develop and properly support and flex the neck. With the musculature of the neck surrounding and attaching to the vertebrae being flexed, it is no surprise that as muscles are removed, maximum flexibility increases. There does not appear to be any group of muscles that specifically affects the total flexibility; though there is a large increase in the maximum dorsal excursions possible in the posterior-most vertebrae after removal of the long, lateral muscles (Fig. 5d), this is likely due to the large amount of tissue that had been removed from those vertebrae (to include the dorsal and ventral muscles). Ventral flexibility is largely limited by the ligamentum elasticum, with extreme excursions possible after the removal of the ligament concurring with Dzemski & Christian (2007) (Fig. 5f). Lateral flexibility is affected by tissue removal in the same way, with overall increases in flexibility. The pattern observed is however much different to that of previous research. Where this study found there is higher flexibility towards the head and middle of the neck, and much lower flexibility at the base (Fig. 6a), the opposite has been presented in prior work (Dzemski & Christian, 2007 (Pg. 707, Fig. 7b), which shows little flexibility at joint 1, uniform flexibility of around 15° from between C2 and C10, and higher flexibility of 20-25° from joints 10 to 18. Discounting differing absolute values due to specimen age or size, as these would be unlikely to change the pattern of flexibility so dramatically, it is likely due to differences in the methods used and observations made. Whilst in the previous study it was stated that “lateral flexibility is significantly reduced if simultaneously flexed dorsally” (Dzemski & Christian, 2007; Pg. 707), during examinations of the ostrich necks the opposite was observed, with only a limited amount of flexibility allowed whilst two vertebrae are dorsoventrally ‘neutral’ (i.e. at 0°). At a certain point dorsal flexion is required to allow for any further lateral excursion, as when the pre-zygapophyses of the posterior vertebrae pass further under the post-zygapophyses of the anterior vertebrae, the body of the posterior vertebra is inevitably lifted upwards (Fig. 15), leading to dorsal flexion. This is due, in part, to the relative width of the pre- and post-zygapophyses, and the angle at which they slope inwards. Where in the more anterior vertebrae the zygapophyses are thinner in relation to length, and the angles are less pronounced, the larger posterior vertebrae have zygapophyses that are relatively much wider and slope dramatically inwards (Fig. 16), This is especially apparent in the posterior-most vertebra, which are naturally inclined towards dorsal flexion (Fig. 5a), and in the case of this study, incapable of ventral excursions with all tissues attached. Inversely, to keep the vertebrae dorsoventrally neutral during larger lateral excursions requires ventral flexion of the anterior vertebrae.

Tissue mass & measurements

The amount of musculature surrounding the vertebrae and joints limits the amount of flexibility in the neck. Whilst osteological stops and ligaments place absolute limits, the amount of musculature around a joint will further limit the maximum flexibility when the animal is alive. There is relatively little difference in the maximum flexibility of the anterior and posterior joints of a neck with little tissue present (Fig. 5e,f), yet there is a much larger difference in one with all musculature intact, with much lower flexibility allowed in the joints towards the base of the neck. As the amount of musculature is much higher in these posterior vertebrae, compared with that of the middle and anterior sections, it is safe to assume that muscle mass has a great deal of influence on the flexibility allowed at the base of the neck, and as this varies not only between species but between individuals, emphasis should be placed on the assumed amount of muscle mass when estimating neck flexibility from fossil specimens. The reduction in flexibility is not caused by the bone itself, as shown by estimates of flexibility from zygapophyseal overlap (Fig. 8). With no tissue present, there is no obvious reduction in the excursions possible in the posterior vertebrae.

Osteological Neutral Pose

Positioning the neck in maximal dorsal flexion allowed when in the ONP does not convey the same pattern (of three sections of flexibility) as that of the neck when manipulated to its actual maximal amount of flexibility. Whilst overall flexibility allowed is much higher in the ONP, there is relatively less flexibility dorsally in the anterior and middle sections of the neck, with the highest flexibilities allowed in the posterior portion, much the opposite of what is implied by maximal flexion. Ventrally there is still little flexibility in the base of the neck compared to the joints anterior to it, but aside from the small amount of flexibility allowed in the joint between the axis and C3, there is no real differentiation between the anterior and middle sections of the neck. Unlike dorsal flexion this is much like the actual pattern observed, however the maximum degree of flexibility is much higher in the ONP. When measuring lateral flexibility there is no clear pattern, whereas with tissues intact there is a higher anterior flexibility, decreasing to very little flexibility at the base of the neck. These findings show that the ONP is not a suitable measure of flexibility of the necks of vertebrates. Whilst a discrepancy between the values for flexibility under the same pattern would allow for adjustments to be made, with the ONP as an over- or under-estimate, the pattern of flexibility across the neck is not conveyed at all aside from in ventral flexion, and as such the ONP does not correctly indicate the flexibility of the cervical column. The amount of overlap between the pre- and post-zygapophyses allowed in the ONP would also appear to be an inappropriate. Where the ONP allows for a minimum of 50% overlap, this minimum is surpassed dorsally between cervicals C7-C10. More interestingly, aside from these three joints the minimum of 50% appears to be an overestimate, with values of around 75-100% overlap more common around the base of the neck. It is also of note that the pattern of minimum overlap allowed follows the same pattern as that of flexibility, with reduced excursions at the anterior of the neck, increased excursions in the middle and the largest amount of overlap at the base of the neck; this means that the minimum amount of overlap is dictated by the flexibility of the joint, and that no one rule for zygapophyseal overlap will convey the flexibility across the whole neck.

When comparing wet, dry and absent cartilage, there is a general increase in flexibility with a reduction in centrum length for each joint, likely due to an increased amount of room for manoeuvrability between said joints. This has direct consequences for assessments of flexibility based on fossil specimens, whether in ONP or through other methods. As the presence of cartilage reduces the amount of flexibility, any attempts to assess flexibility through dry bone alone must be overestimates due to an under-represented total centrum size. However, the length of the neck decreases when all centra are placed in contact with each other. This indicates that the centra of the neck are not in constant contact with each other, and there is a certain amount of space between vertebrae within the synovial capsules. This is best illustrated by comparing the neck in sub-maximum flexibility prior to dissection, and the neck skeleton articulated to fit the maximum flexibility of the neck with all tissue intact, but with the centra touching (Fig. 17). The ONP does not allow for these deviations, keeping a constant (and minimum) gap between two centra. As there is this room for manoeuvrability, it is possible that the same amount of flexibility can be obtained with a reduced deviation from neutral zygapophyseal overlap (Fig.

18). This suggests that the ONP could also lead to underestimates of potential flexibility. Coupled with the differences in flexibility allowed due to the size/presence of cartilage, these findings have huge bearings on future estimates of flexibility. To correctly estimate flexibility of the necks of extinct animals from fossil material alone would also require estimates of the size of any cartilage, along with an estimate of the maximum space allowed in the synovial capsules between adjacent vertebrae.

Proportions of attachment sites

The ansa costotransversaria of the sauropods follows a much steeper increase in proportional size than that of the ostrich. With large increases in proportional size towards the base of the neck, it can be expected that the size of the m. ascendens cervicalis increased in size at a much higher rate towards the base of the neck than it does in the ostrich. The m. ascendens cervicalis is utilised in dorsiflexion of the neck. It is likely due to the extremely large size, and the increased moment arm from the head to the posterior cervicals, that the increase in muscle size is needed in order to lift and stabilise the neck. The change in size is also apparent between sauropods, with the more gracile *Diplodocus* having a lighter neck and a smaller attachment site size, whereas the more robust *Apatosaurus* neck would be much heavier, and as such requiring a larger muscle to accommodate the extra weight.

The processus spinosus of the sauropods also increases in proportional size towards the base of the neck, but not to the extent of the ansa costotransversaria. From this attachment site, the m. long. col. dors. and the mm. interspinales originate. M. long. col. dors. is involved not only in dorsiflexion, but in ventriflexion, acting as a support muscle to keep the neck stable, and is used when retracting the neck dorsally from a ventral pose. The need for an increased muscle size here is probably the same as for the ansa costotransversaria. The mm. interspinales are short inter-vertebral muscles running from the spinuous process of one vertebra to the same site on the immediately adjacent vertebra. They are likely to aid dorsiflexion, but also place a limit on ventral flexibility. These muscles are not well defined in extant birds, and it is likely that the change in size of the processus spinosus in sauropods is mostly due to a change in size of the m. long. col. dors.

The tuberculum ansa shows a general trend towards increased size towards the base of the neck in all species, though again with the sauropods having larger proportional attachment sites than that of the ostrich. The m. flex. col. lat and the mm. intertrans. both originate from the tuberculum ansa. The mm. intertrans. is utilised in lateral flexion of the neck as there are no long laterally-flexing muscles in bird necks. Whether this is the case for sauropods is unknown, as it would require novel musculature not present in extant avians. The m. flex. col. lat. aids ventral excursions of the neck. The large increase in size in the tuberculum ansa of sauropods may be due to the higher position of the shoulders, allowing ventriflexion of the neck so that the head reaches the ground.

The processus costalis of the *Apatosaurus* is relatively much larger than that of the ostrich across the whole neck, whilst similar in size between vertebra. The m. flex. col. med. originates from this site, and as with the m. flex. col. lat. is involved in ventral flexion of the cervical column. As the processus caroticus does not have an equivalent attachment site in sauropods, and this is the site where the long ventral muscles originate in birds, it may be the case that the increased size of the intervertebral ventrally flexing muscles is to compensate for a lack of these longer muscles. The processus caroticus in birds also shows a linear increase in relative size over the length of the bird neck, allowing ventral flexion from a resting raised neck. The lack of this attachment site and its corresponding muscles could be construed as an indicator of a more horizontal neck posture compared to birds, as it would not require the musculature to bring the anterior portion of the neck down to feed on low plants or drink from water sources.

The relative attachment site sizes of the crista transverso-obliqua/ spinopostzygapophyseal lamina, and the torus dorsalis, both show little signs of any trend in proportional size in any species, and the species analysed do not differ greatly enough to warrant further examination.

## Implications for Sauropods

With regards to overall flexibility of the neck, it has been shown that the ONP could potentially lead to either underestimates or overestimates of flexibility; as it overestimates that of the ostrich whilst not accounting for any gaps in the centra we can assume it is a general overestimate. This would decrease the flexibility of the sauropod neck, and therefore decrease the potential range of the feeding envelope over which it was possible for them to feed. This would facilitate even greater niche partitioning than previously suggested in the literature (the ONP gives the lowest estimate in feeding envelope size (Stevens & Parrish, 2005b)). This reduction would potentially bring the feeding envelope of sauropods with necks that certainly had more vertical neutral postures (at least at the base of the neck), such as the Macronarian *Brachiosaurus*, out of range of potential water sources. However, this is not a paradoxical scenario. Barring a novel structure such as an elephant’s trunk, it is entirely possible that the sauropods were capable of kneeling to bring their heads closer to the water level. Whilst the obvious example of the giraffe splaying its legs would not apply to the much more robust sauropods, it is important to remember that this behaviour in the giraffe is necessary due to the elongate metapodials. This elongation is not exhibited in sauropods, with the knee joint much closer to the centre of the limb as a whole, allowing the knees to bend and bring the body downwards whilst keeping the manüs directly below the body to continue to support the weight of the animal.

A decrease in flexibility does however put limitations on the resting posture of the neck, in particular suggestions of a swan-like 'S' shaped posture. A higher head height coupled with a lower flexibility would prevent the head reaching water sources to drink, with the ability to bend the knees only adding a certain amount leeway. Of course it is conceivable that the 'neutral' (i.e. posture that uses the least energy to maintain) posture of the neck is much lower than that of the posture whilst resting, however it would be energetically inefficient to constantly hold the neck close to dorsally flexed the majority of the time. Therefore it is likely that given a decreased feeding envelope, an 'S' shaped neck would be impractical. However it is entirely possible for the neck to have been held in a posture raised slightly above horizontal (Christian & Dzemski, 2007; Dzemski & Christian, 2007; Christian & Dzemski, 2011). The lack of an attachment site that is homologous to the avian processus caroticus suggests one of two things: either there was a novel attachment site that has yet to be identified in the sauropod neck that long, ventral muscles originated from, or these muscles were not present. Without these long ventral muscles, ventral excursions would be limited, implying that a swan-shaped neck would not be possible, as the animal would not be able to lower its head down sufficiently.

With regards to flexibility of individual joints of the neck, it is clear that the sauropods have relatively more mass to restrict flexibility at the posterior portion of the neck compared to the ostrich. Whilst the ostrich has very little flexibility at the base of its neck, a reduction in this already small range would seem a hindrance. However, as the length of the neck is much longer in the sauropod than it is in the ostrich, a smaller degree of flexibility would allow for a much larger change in height at the anterior end of the neck. This increase in muscle mass is most likely necessitated by the need to compensate for the increased moment arm produced by a much longer and heavier neck. The muscles that are implied to have an increased relative mass in sauropods include (but are not limited to) those that aid dorsiflexion and stability, which in the ostrich are much more pronounced at the base of the neck. The only other study to deal with flexibility estimates for the individual joints of the neck is Dzemski & Christian (2007). It was proposed that dorsal flexibility was limited by bone, and that ventral excursions were limited to a minimum of 30% zygapophyseal overlap. The results presented here assert that these limitations are demonstrably false. As flexibility is increased through the removal of muscles, bone cannot be the limiting factor in dorsal flexibility. In addition, zygapophyseal overlap in the ostrich is at minimum around 60%, following a pattern where more flexible joints show lower overlaps and vice versa. It has also been shown that allowing for a fluctuating gap between the centra allows a higher amount of flexibility with the same zygapophyseal overlap. The evidence suggests that using percentage overlap of dry bone is not an appropriate measure of flexibility. It is of note that the ventral flexibilities proposed in Dzemski & Christian (2007) (Pg. 709, Fig. 10) contained two large spikes in flexion capabilities at the 8th and 15th joints (accompanied by large drops in dorsal flexion). When viewing the estimated ventral flexibilities of an ostrich in the ONP, which again is based on zygapophyseal overlap, the same comparatively high flexibilities over individual joints can be seen in the 5th, 8th and 11th joints, again accompanied by a reduction in dorsal flexion in comparison to the prior vertebrae. This is in conflict with the pattern of actual maximal flexibility which is a much smoother trend divided into three broad sections. It is much more likely that the neck of the sauropods would transition in this smooth fashion, as such large variations in flexibility would require a considerable amount more muscle localised around individual vertebra to accommodate this increase in flexibility; this would be required to bring the joint back up to a more neutral posture. This is not shown in the *Diplodocus* (the same specimen used by Dzemski & Christian, 2007), with attachment site values for the dorsal and ventral muscles showing no obvious decrease or increase in relative size around this joint. However, the processus costalis was not present in the vertebrae of this specimen. In the *Apatosaurus* the cervical ribs were present, and there is a pronounced increase in relative size at the 8th vertebrae, where a large amount of muscle devoted to ventral flexion would attach. Though this large attachment site is present, it is unlikely that the neck of the sauropod contained a much larger amount of mass concentrated around the middle of the neck.

## Conclusions

- The ostrich neck can be divided into three sections of varying flexibility; a slightly flexible anterior section, a very flexible middle section, and a stiff posterior section.
- The muscles of the neck are what place limits on flexibility, as removal of the muscles leads to higher maximum flexibility. Therefore muscle mass needs to be taken into account in any predictions of flexibility.

- Zygapophyseal overlap of bone does not indicate flexibility. Sections of the neck with lower flexibilities show more overlap, and vice versa. Therefore the Osteological Neutral Pose is inappropriate as a measure of flexibility of the neck.
- The size of cartilage, as well as its presence, affects potential flexibility. This, and the fact that the inter-vertebral spaces are not kept to an absolute minimum at all times, mean that any further work requires the space between two centra to be taken into account to come to a meaningful conclusion.
- If the Osteological Neutral Pose affects estimates of sauropods in the same way it does the ostrich (a general overestimate), sauropod neck flexibility is lower than previously imagined. Therefore the range of their feeding envelopes would be much smaller than prior estimates.
- Limited flexibility would prevent more vertical, ‘S’-shaped necks due to an inability to reach water sources.

**Further Work**

It is important to note that as the ONP both underestimates and overestimates the flexibility of the joints of the neck, it is entirely possible that for some species, such as the ostrich presented here, it overestimates flexibility of the whole neck, and for some species it underestimates this. More studies into the flexibility of extant animal necks would lead to a more definitive answer to this. As the original DinoMorph and its successive revisions are the only current computer models of sauropod neck flexibility, they are valuable in that their results can be used to base comparisons of actual flexibilities and those provided by the ONP. Definite candidates for further work in this area include the rhea and other extant avians with elongate necks, and mammals such as the giraffe and the camel. Though not exhibiting elongate necks, crocodylians are in dire need of assessment to properly bracket the sauropods.

Should further work be completed on attachment site sizes, the rhea is again an ideal candidate for avian musculature, especially as it is the only extant bird that exhibits bifid neural spines (Tsuihiji, 2004). Although many recovered sauropod cervical series are subject to deformation, poor preservation and loss of one or more vertebrae, there are still well preserved representatives available. Measurements of the attachment sites of macronarian sauropods such as *Brachiosaurus* or *Camarasuarus* would prove to be the most informative due to their dramatically different morphology compared to the diplodocids studied here.

## Table Legends

**Table 1**. Muscle attachment site abbreviations used in figures

**Table 2**. Origins and insertions of the cervical musculature of *Struthio camelus* (the ostrich). Muscles appear in the order removed in this study. Modified from Wedel & Sanders, 2002.

**Table 3**. Mass measurements of the muscle groups of the neck of *Struthio camelus* (the ostrich). Also presented are groups as a percentage of the total muscle mass of the neck, and as a percentage of the total mass of the neck (Dorsal: M. biv. cerv., m. long. col. dors., m. asc. cerv.; Ventral: m. flex. col. med., m. long. col. ven.; Lateral: M. flex. col. lat.; Single-segment: Mm. intercristales, mm. interspinales, mm. intertrans.).

**Table 4**. Length measurements of the centra of the neck of *Struthio camelus* (the ostrich). Measurements were taken whilst cartilage was wet after boiling off tissue; after 4 days of drying; after removal of the cartilage from the vertebra. All measurements in cm.

## Figure Legends

**Figure 1**. Measuring inter-vertebral flexibility of *Struthio camellus* with a medical goniometer. (a) Measuring flexion of the neck with muscles intact. (b) Measuring flexion of the cleaned vertebra using adjustable clamps.

**Figure 2**. Mid-cervical vertebrae of *Struthio camelus* (a, b) and *Apatosaurus louisae* (c, d), with muscle attachment sites labelled. Vertebrae illustrated in left lateral (a, c) and anterior (b, d) views.

**Figure 3**. The neck of *Struthio camelus*. Annotated to show the muscular bellies and tendons of m. biventer cervicis. Scale bar = 10cm.

**Figure 4**. The neck of *Struthio camelus*, annotated to show the muscles mm. intertransversii, m. ascendens cervicalis, m. longus colli dorsalis, and the location of the muscle attachment sites torus dorsalis and ansa costotransversaria. Scale bar = 10cm.

**Figure 5**. Measurements of dorsoventral flexibility of the neck joints of *Struthio camelus* through stages of cumulative tissue removal. (a) All tissues present. (b) Long dorsal muscles removed. (c) Long ventral muscles removed. (d) Long lateral muscles removed. (e) Single-segment muscles removed. (f) Ligamentum elasticum removed. ((a) n=3; (b-e) n=2; (f) n=1).

**Figure 6**. Measurements of lateral flexibility of the neck joints of *Struthio camelus* through stages of cumulative tissue removal. (a) All tissues present. (b) Long dorsal muscles removed. (c) Long ventral muscles removed. (d) Long lateral muscles removed. (e) Single-segment muscles removed. (f) Ligamentum elasticum removed. ((a) n=3; (b-e) n=2; (f) n=1).

**Figure 7**. Mass measurements of the neck of *Strutho camelus*, after the neck was separated at each individual joint. (a) Mass of each cervical vertebra and the tissue surrounding it. (b) Estimated tissue mass around each inter-vertebral joint.

**Figure 8**. Measurements of flexibility of the neck skeleton of *Struthio camelus* when limited to a minimum of 50% zygapophyseal overlap, to conform with the osteological neutral pose (Stevens & Parrish, 1999). (a) Dorsoventral flexibility. (b) Lateral flexibility. (n=3).

**Figure 9**. The degree of dorsal flexion at each joint when the cervical vertebrae of *Struthio camelus* are articulated in the osteological neutral pose (100% zygapophyseal overlap) (Stevens & Parrish, 1999). (n=3) .Figure 10. Neck skeleton (C3-C17) of *Struthio camelus* articulated to show (a) maximum dorsal flexibility; (b) neutral position; (c) maximum ventral flexibility allowed by the osteological neutral pose (Stevens & Parrish, 1999). Scale bar = 10cm.

**Figure 11**. Maximum dorsoventral flexibility of the neck skeleton of *Struthio camelus* allowed by the osteological neutral pose (minimum 50% zygapophyseal overlap), (a) whilst the cartilage of the vertebra was wet after boiling off tissue; (b) after drying for 4 days; (c) after removal of the cartilage.

**Figure 12**. Estimated zygapophyseal overlap of the cervical vertebra of *Struthio camelus* whilst the complete neck with all tissue intact is in maximum dorsal and maximum ventral flexion. (n=3).

**Figure 13**. Proportional size of the attachment sites of the cervical muscles along the necks of *Struthio camelus*, *Diplodocus carnegii* and *Apatosaurus ajax*, in relation to length of the respective centrum of each vertebra. (a) Ansa costotransversaria; (b) processus spinosus; (c) crista transverso-obliqua (*S. camelus*) and spinopostzygapophyseal lamina (*D. carnegii* & *A. ajax*); (d) torus dorsalis; (e) tubercula ansae; (f) processus costalis; (g) cristae lateralis; (h) processus caroticus.

**Figure 14**. Proportional size of the attachment sites of the cervical muscles along the necks of *Struthio camelus*, *Diplodocus carnegii* and *Apatosaurus ajax*, in relation to the total neck length. (a) Ansa costotransversaria; (b) processus spinosus; (c) crista transverso-obliqua (*S. camelus*) and spinopostzygapophyseal lamina (*D. carnegii* & *A. ajax*); (d) torus dorsalis; (e) tubercula ansae; (f) processus costalis; (g) cristae lateralis; (h) processus caroticus.

**Figure 15**. The effect of lateral flexion on dorsoventral flexion in the posterior cervical vertebrae of *Struthio camelus*. (a, c) C15 and C16 with no lateral flexion, and flexed ventrally to reach a dorsoventral angle of 0° (see zygapophyseal overlap (a)). (b, d) C15 and C16 flexed laterally, forcing dorsal flexion.

**Figure 16**. Pre- and post-zygapophyses of the cervical vertebrae of *Struthio camelus*. C10 (a) pre-zygapophyses; (b) post-zygapophyses. C15 (c) pre-zygapophyses; (d) post-zygapophyses.

**Figure 17**. The effect of inter-vertebral space on zygapophyseal overlap in the neck of *Struthio camelus*. (a) C11 and C12 in 20° dorsiflexion with no space between centra, with zygapophyseal overlap shown (c). (b) C11 and C12 in 20° dorsiflexion with 0.2cm gap between centra, with increased overlap of zygapophyses (d). Scale bars = 2cm.

**Figure 18**.The effect of inter-vertebral space on overall flexibility of the neck of *Struthio camelus*. (a) neck with all tissues intact in sub-maximal dorsiflexion. (b) the same neck cleaned of all tissue, articulated to match the maximum dorsal flexibility of each joint, with all centra touching. Scale bars = 10cm.

**Tables**

**Table 1.**

| Abbreviation | Attachment site |
|---|---|
| **act** | Ansa costotransversaria |
| **cl** | Crista lateralis |
| **cto** | Crista transverso-obliqua |
| **pca** | Processus caroticus |
| **pco** | Processus costalis |
| **psp** | Processus spinosus |
| **spol** | Spinopostzygapophyseal lamina |
| **ta** | Tuberculum ansa |
| **td** | Torus dorsalis |

**Table 2.**

| Table 2.Muscle | Origin | Insertion |
|---|---|---|
| **M. biventer cervicis** | Processus spinosus of the posterior cervical/anterior thoracic vertebrae | Parietals |
| **M. longus colli dorsalis** | Processus spinosus | Torus dorsalis |
| **M. ascendens cervicalis** | Ansa costotransversaria | Torus dorsalis |
| **M. flexor colli medialis** | Processus caroticus<br>Processus costalis | Processus ventralis corporis<br>Processus costalis |
| **M. longus colli ventralis** | Processus caroticus<br>Processus ventralis corporis | Processus costalis |
| **M. flexor colli lateralis** | Tubercula ansae<br>Cristae laterales | Processus costalis |
| **Mm. intercristales** | Crista transverso-obliqua | Crista transverso-obliqua |
| **Mm. insterspinales** | Processus spinosus | Processus spinosus |
| **Mm. intertransversarii** | Tubercula ansae<br>Cristae laterales | Tubercula ansae<br>Cristae laterales |

**Table 3.**

| Muscle group | Mass (g) | % of total muscle mass | % of total neck mass |
|---|---|---|---|
| **Dorsal** | **433** | **40.53** | **22.26** |
| *Of which m. biv. cerv.* | *253* | *23.69* | *13.01* |
| **Ventral** | **176** | **16.47** | **9.05** |
| **Lateral** | **184** | **17.22** | **9.46** |
| **Single-segment** | **275** | **25.74** | **14.14** |
| *Of which dorsal* | *104* | *9.74* | *5.35* |
| *Of which mm. intertans.* | *171* | *16.01* | *8.79* |
| **Total** | **1068** | | |

**Table 4.**

| Vertebra | Wet | Dry | Removed |
|---|---|---|---|
| **C3** | 4.3 | 4 | 3.7 |
| **C4** | 4.85 | 4.7 | 4.5 |
| **C5** | 5.55 | 5.2 | 5.2 |
| **C6** | 5.4 | 5.3 | 4.9 |
| **C7** | 5.8 | 5.5 | 5.35 |
| **C8** | 5.9 | 5.8 | 5.5 |
| **C9** | 6.1 | 6 | 5.8 |
| **C10** | 6.2 | 6.15 | 6.1 |
| **C11** | 6.5 | 6.5 | 6.3 |
| **C12** | 6.8 | 6.7 | 6.45 |
| **C13** | 7.05 | 7 | 6.7 |
| **C14** | 7.1 | 7 | 6.9 |
| **C15** | 7.6 | 7.3 | 7 |
| **C16** | 7.4 | 7.2 | 7 |
| **Total** | 86.55 | 84.35 | 81.4 |

**Figures**

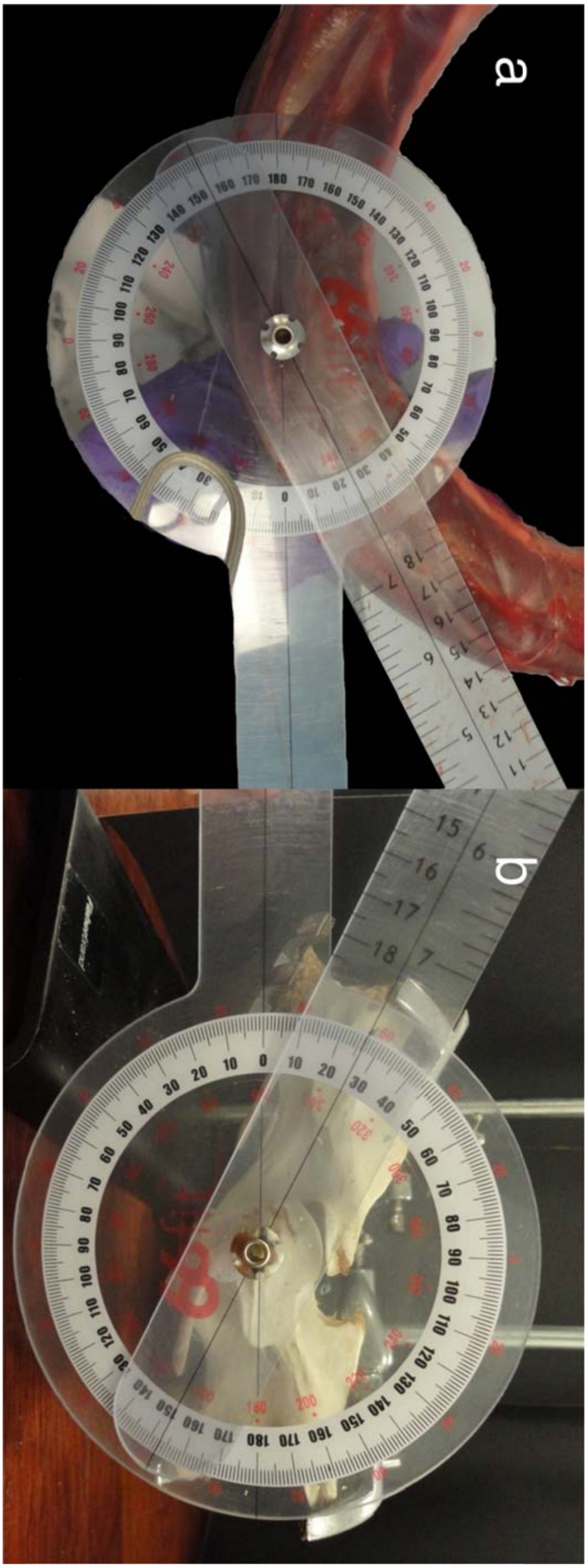

**Figure 1.**

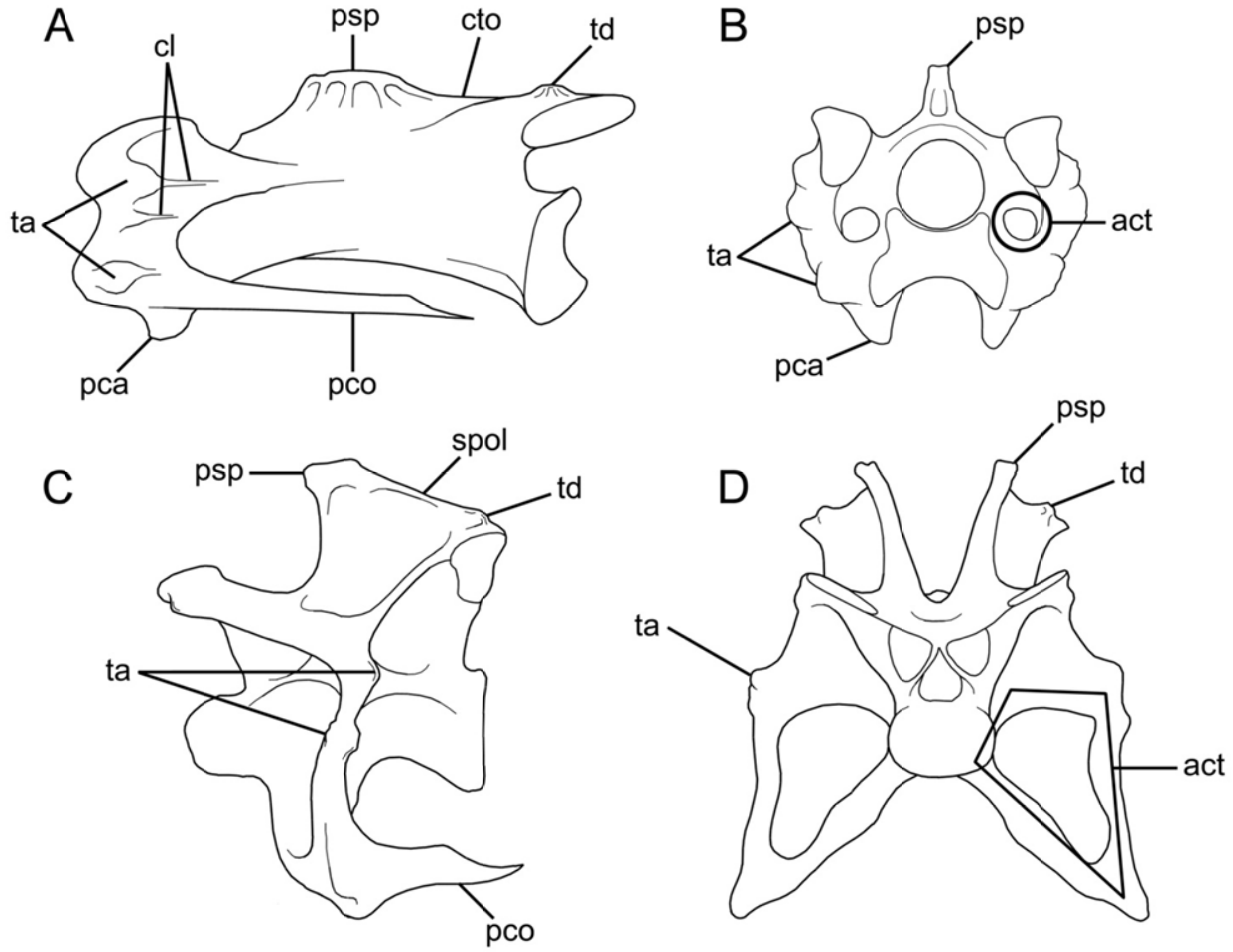

**Figure 2.**

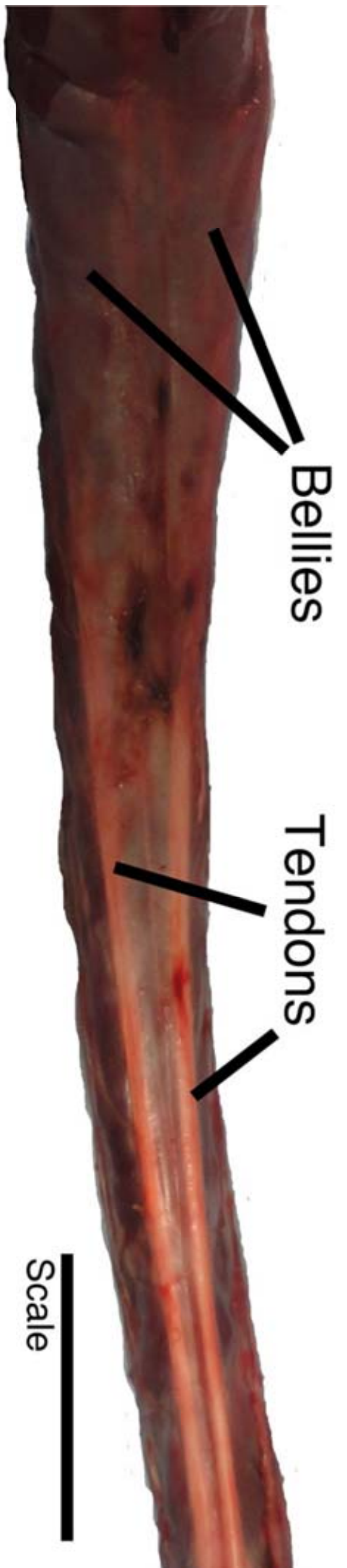

**Figure 3.**

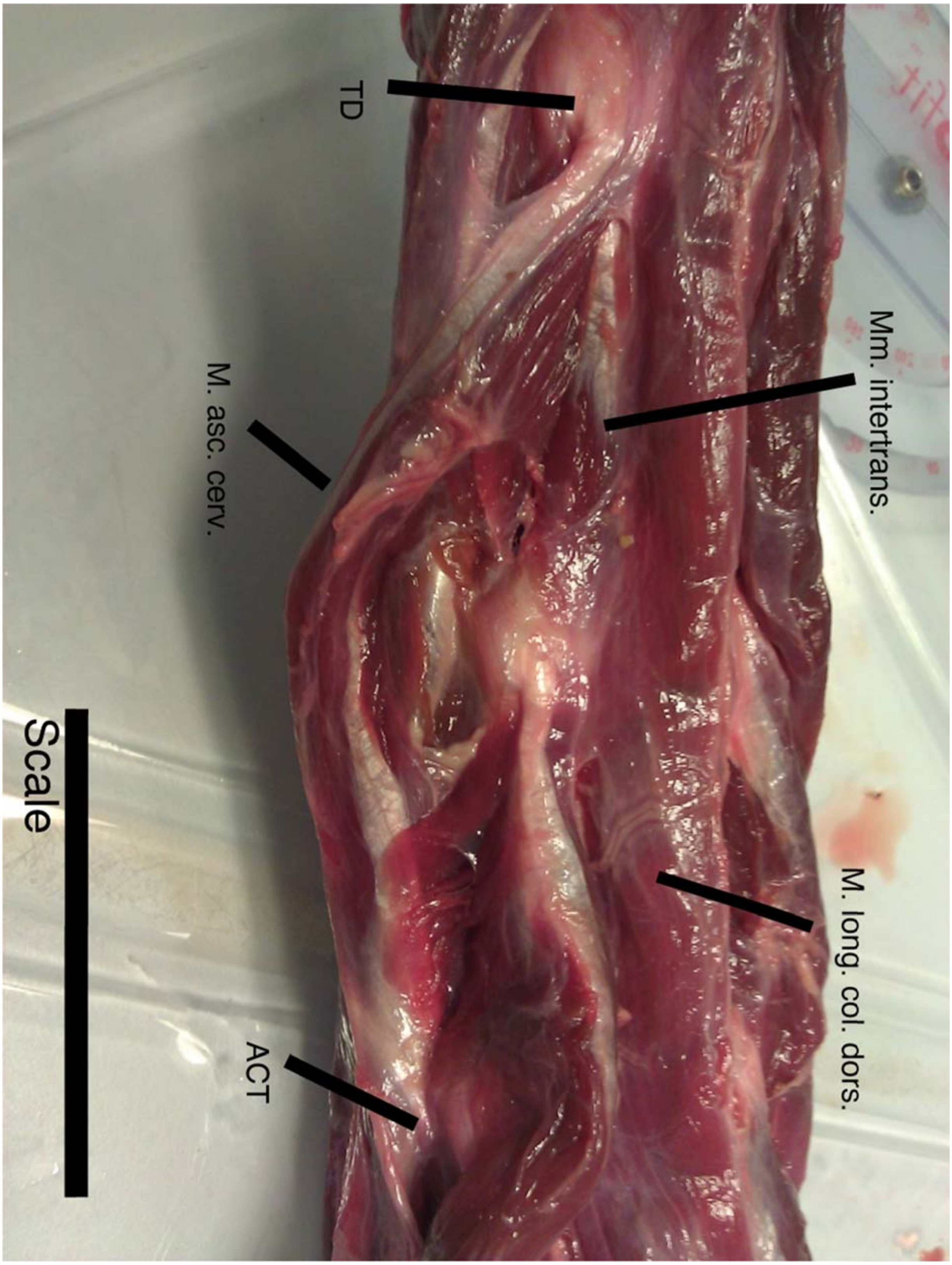

**Figure 4.**

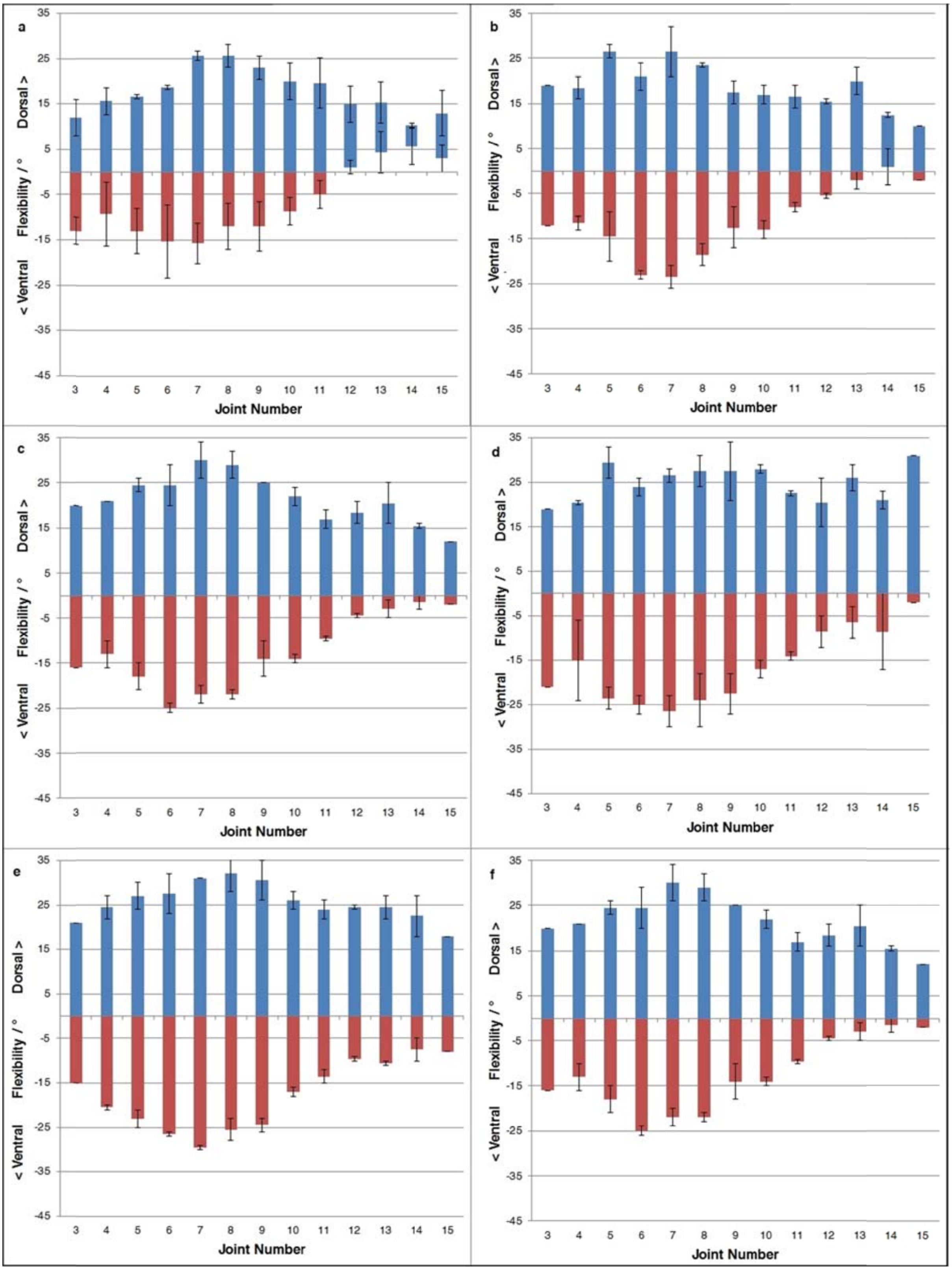

**Figure 5.**

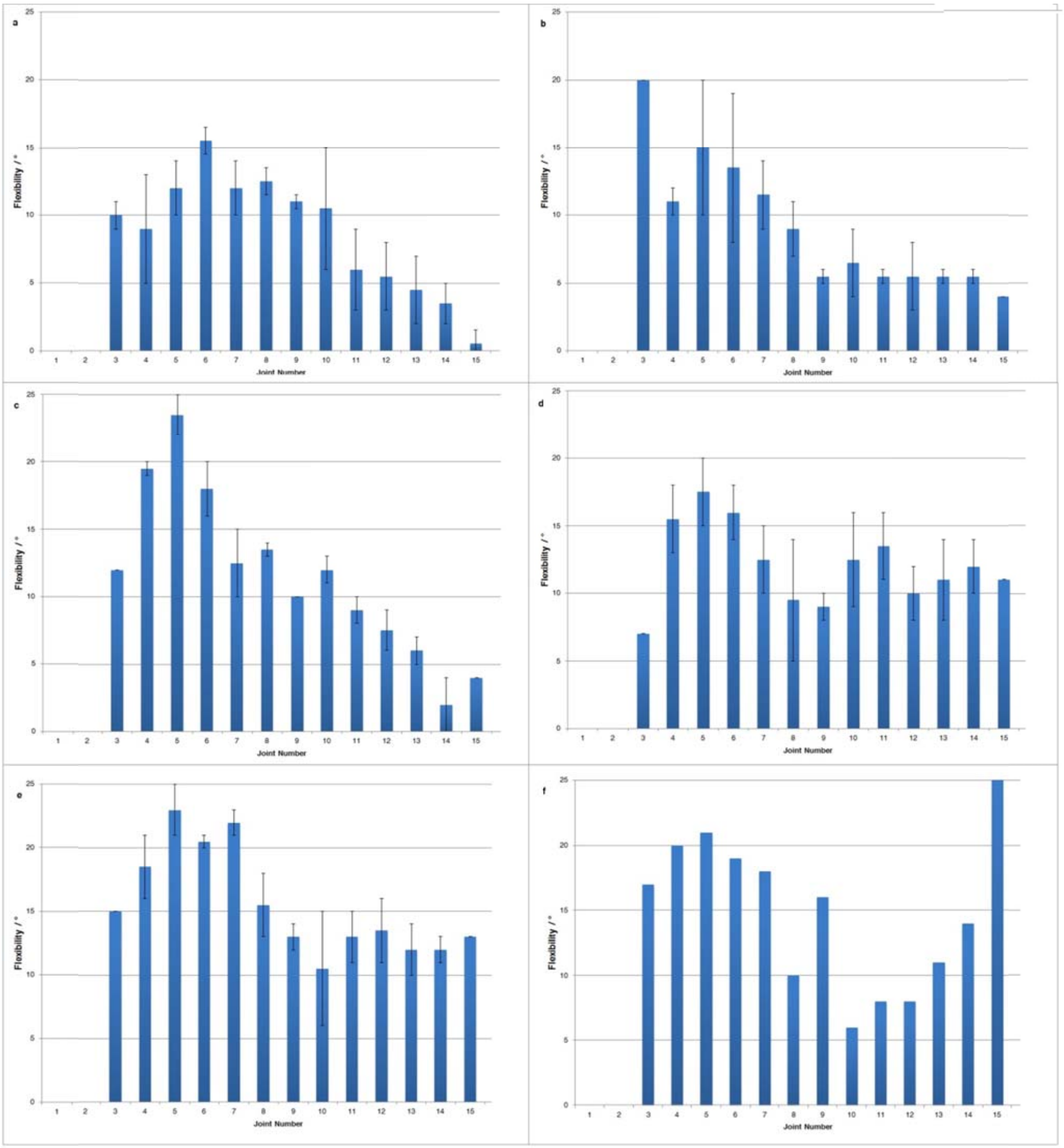

**Figure 6.**

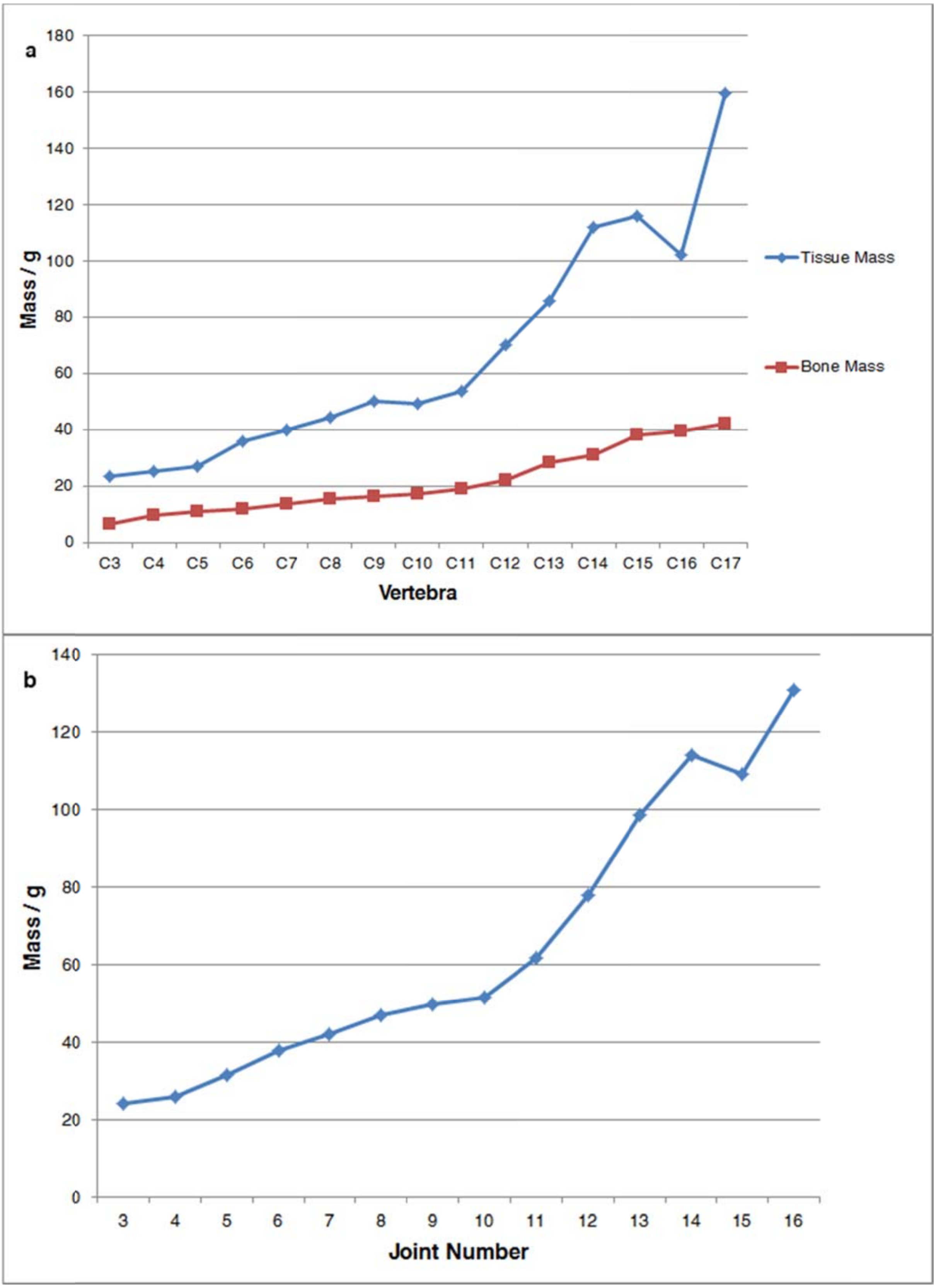

**Figure 7.**

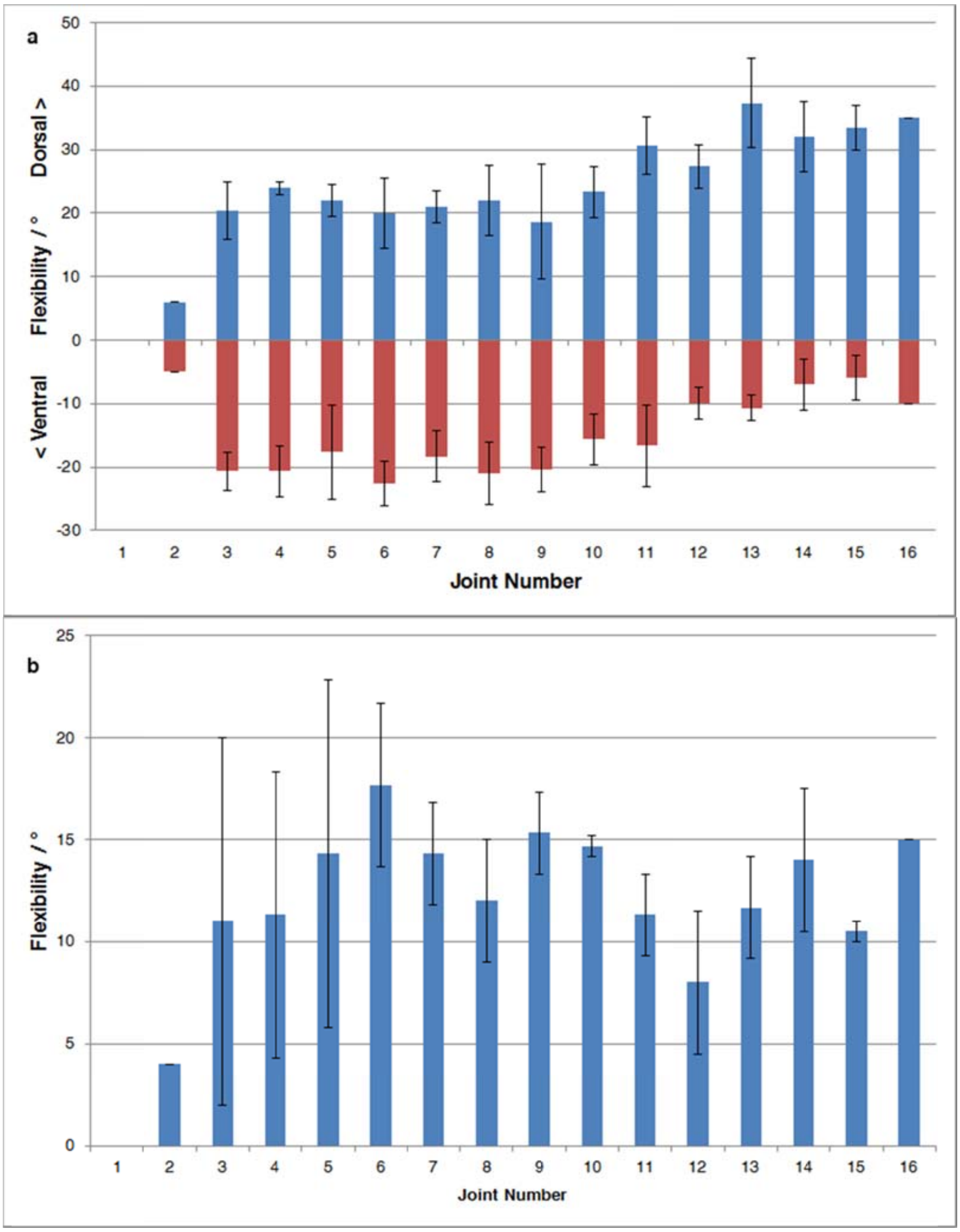
a
50
40
30
20
10
0
-10
-20
-30
Dorsal >
Flexibility / °
< Ventral
1 2 3 4 5 6 7 8 9 10 11 12 13 14 15 16
Joint Number
b
25
20
15
10
5
0
Flexibility / °
1 2 3 4 5 6 7 8 9 10 11 12 13 14 15 16
Joint Number

**Figure 8.**

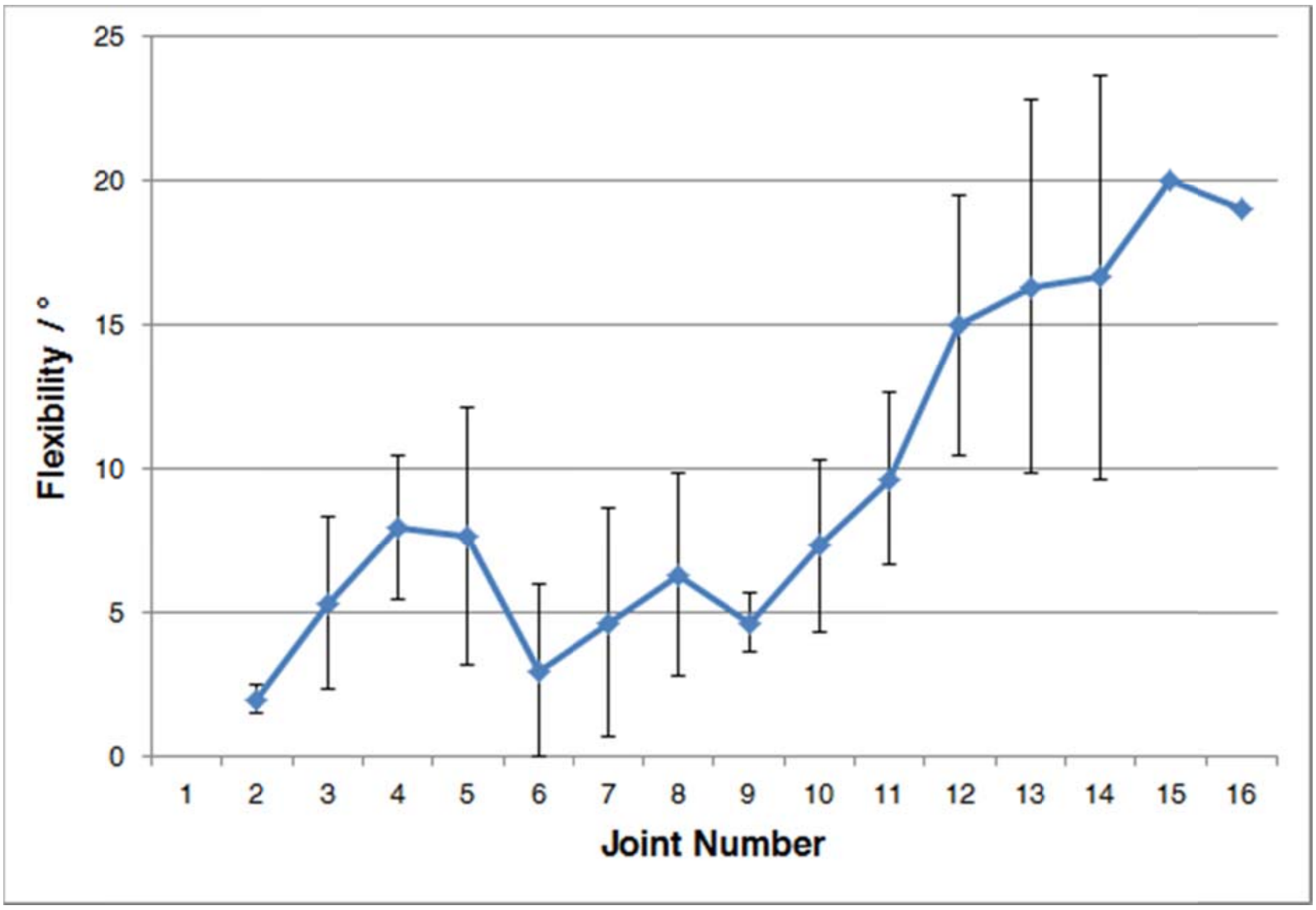

**Figure 9.**

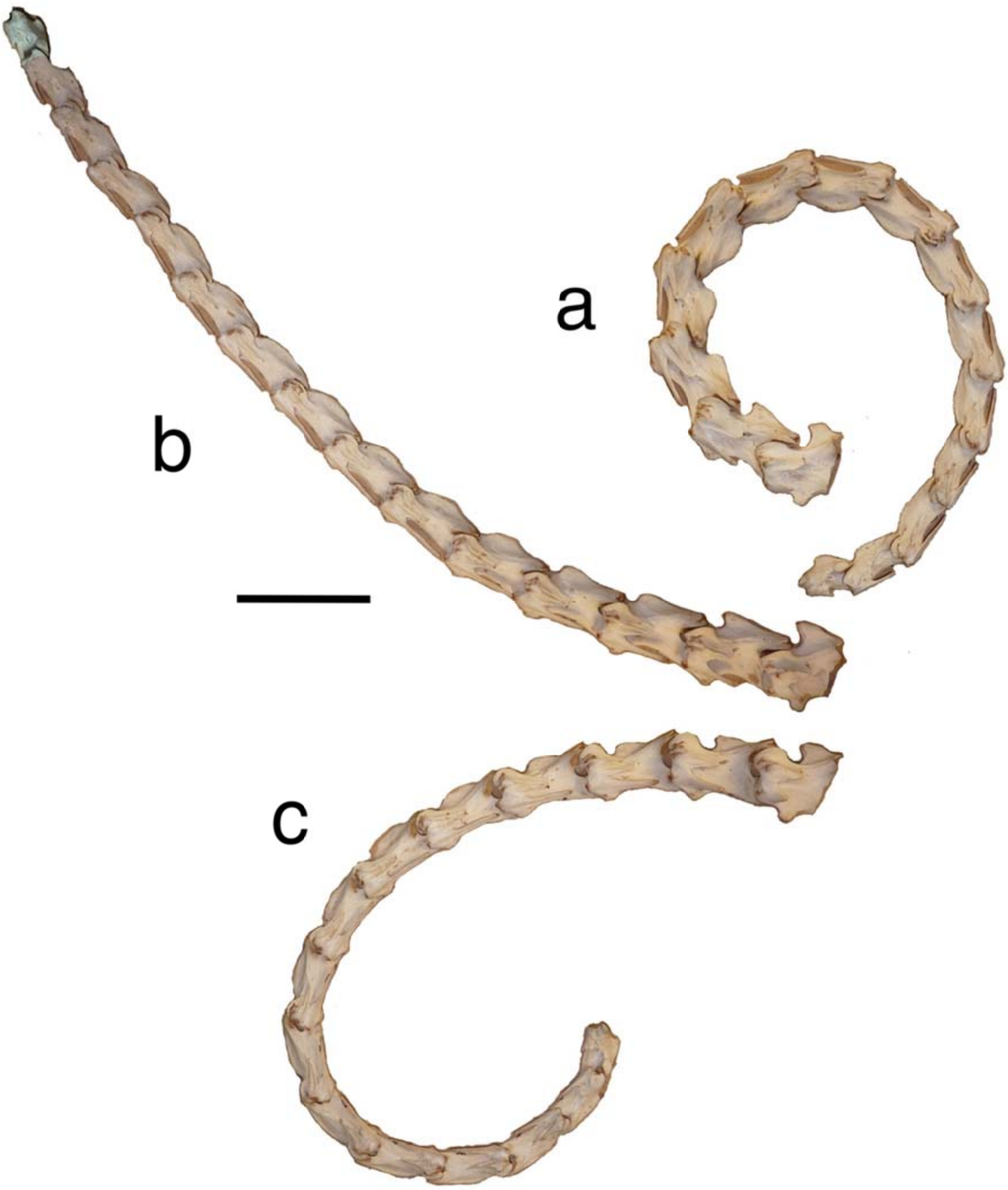

**Figure 10.**

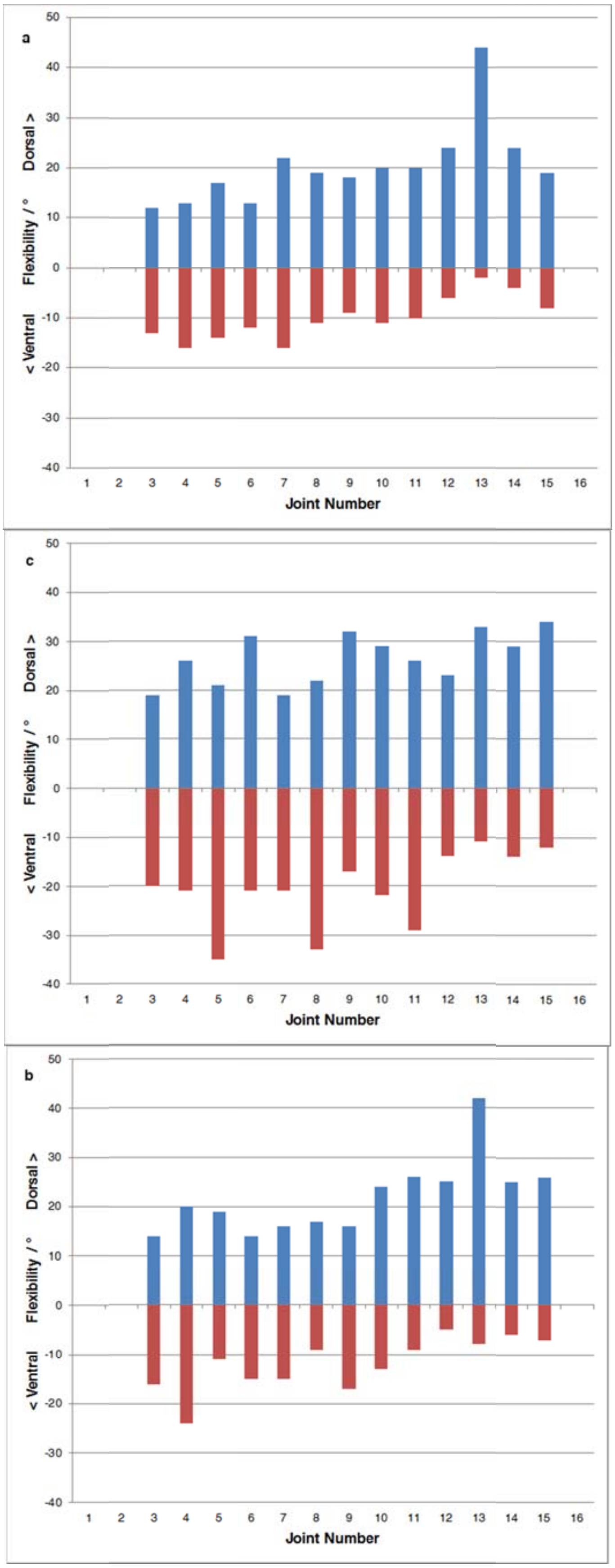

**Figure 11.**

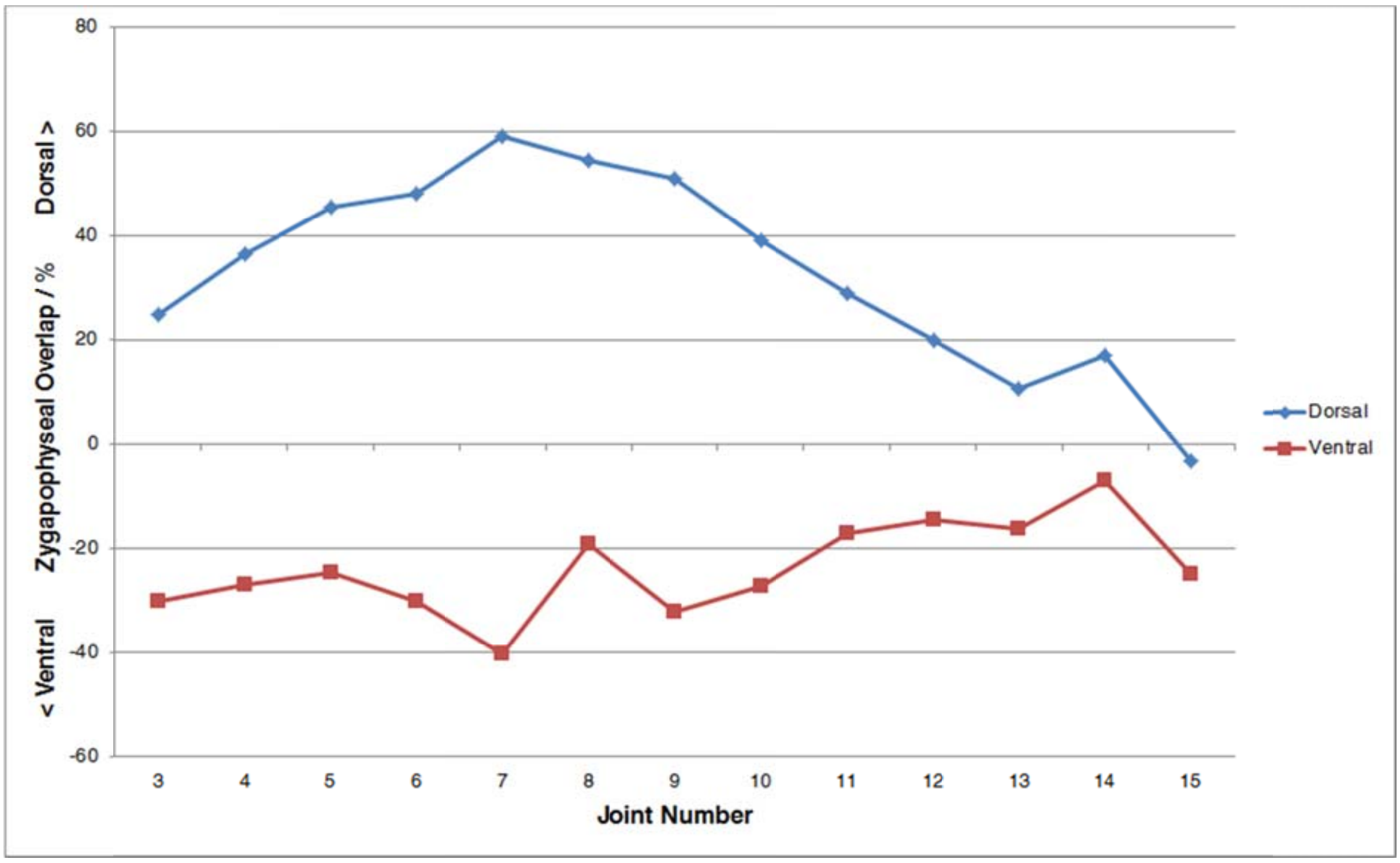
80
60
40
20
0
-20
-40
-60
< Ventral Zygapophyseal Overlap / % Dorsal >
3
4
5
6
7
8
9
10
11
12
13
14
15
Joint Number
Dorsal
Ventral

Figure 12.

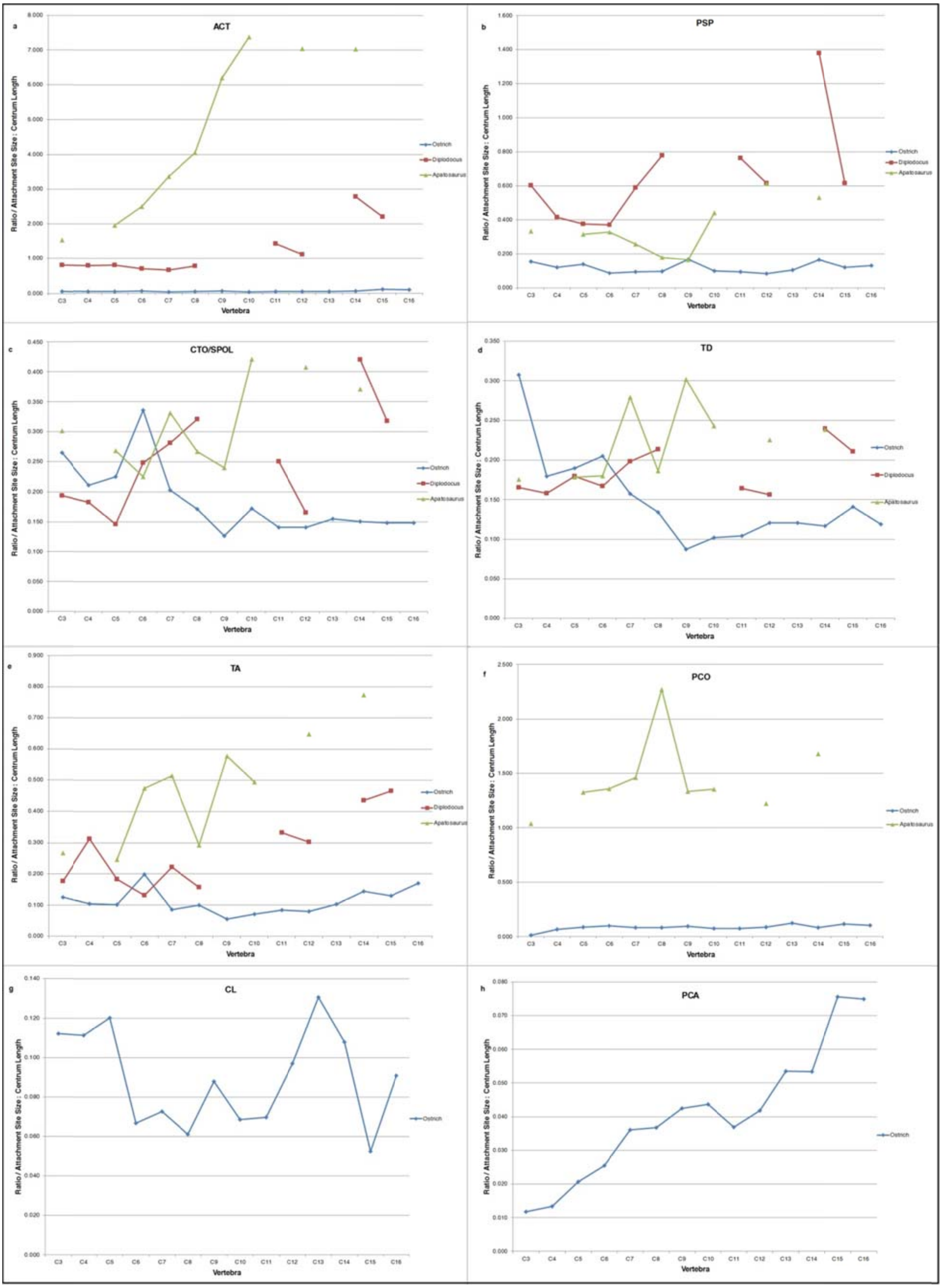
a
ACT
b
PSP
c
CTO/SPOL
d
TD
e
TA
f
PCO
g
CL
h
PCA
Ratio / Attachment Site Size : Centrum Length
Vertebra
Ostrich
Diplodocus
Apatosaurus

**Figure 13.**

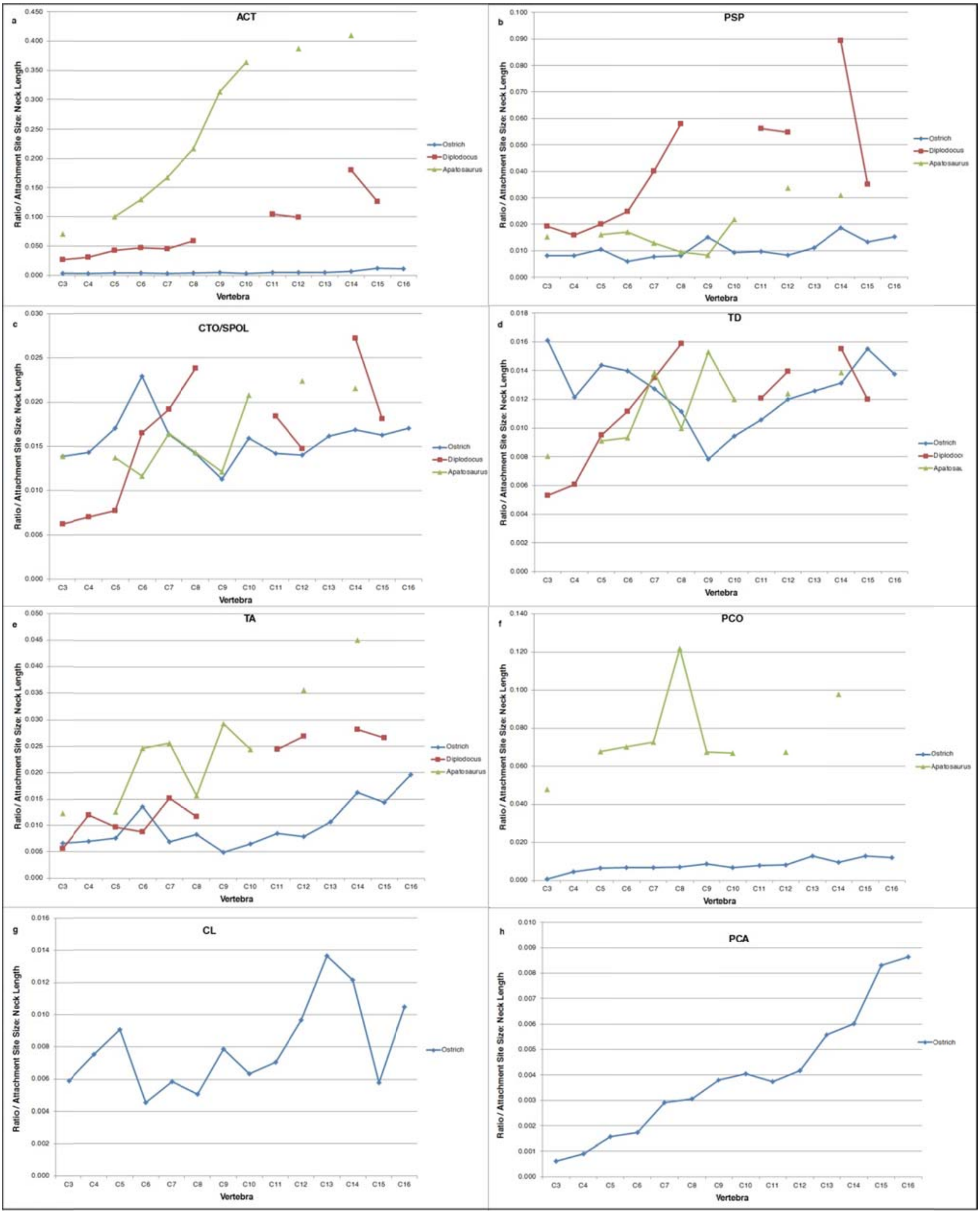

a
ACT
b
PSP
c
CTO/SPOL
d
TD
e
TA
f
PCO
g
CL
h
PCA
Ratio / Attachment Site Size: Neck Length
Vertebra
C3 C4 C5 C6 C7 C8 C9 C10 C11 C12 C13 C14 C15 C16
Ostrich
Diplodocus
Apatosaurus

**Figure 14.**

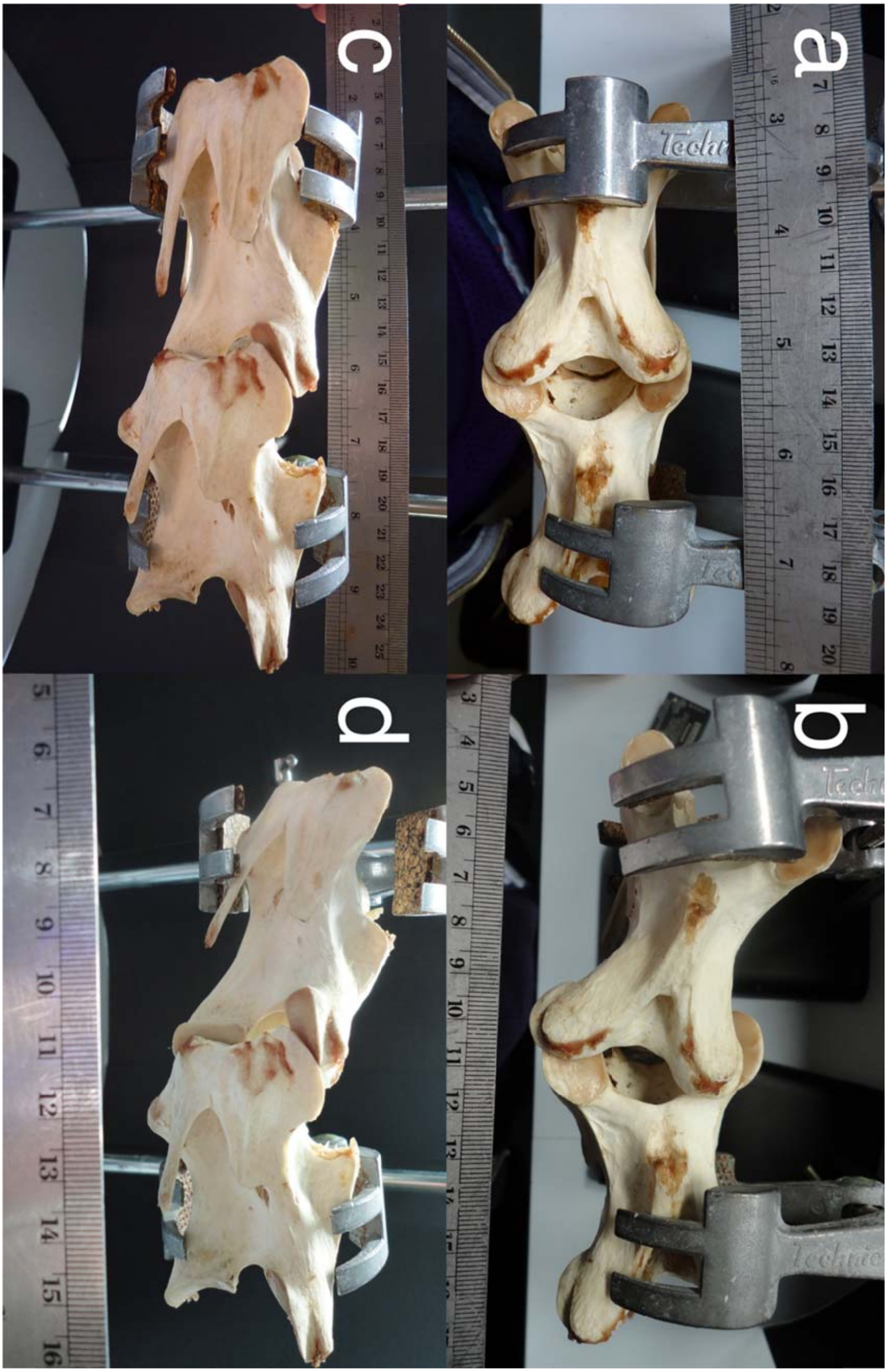

**Figure 15.**

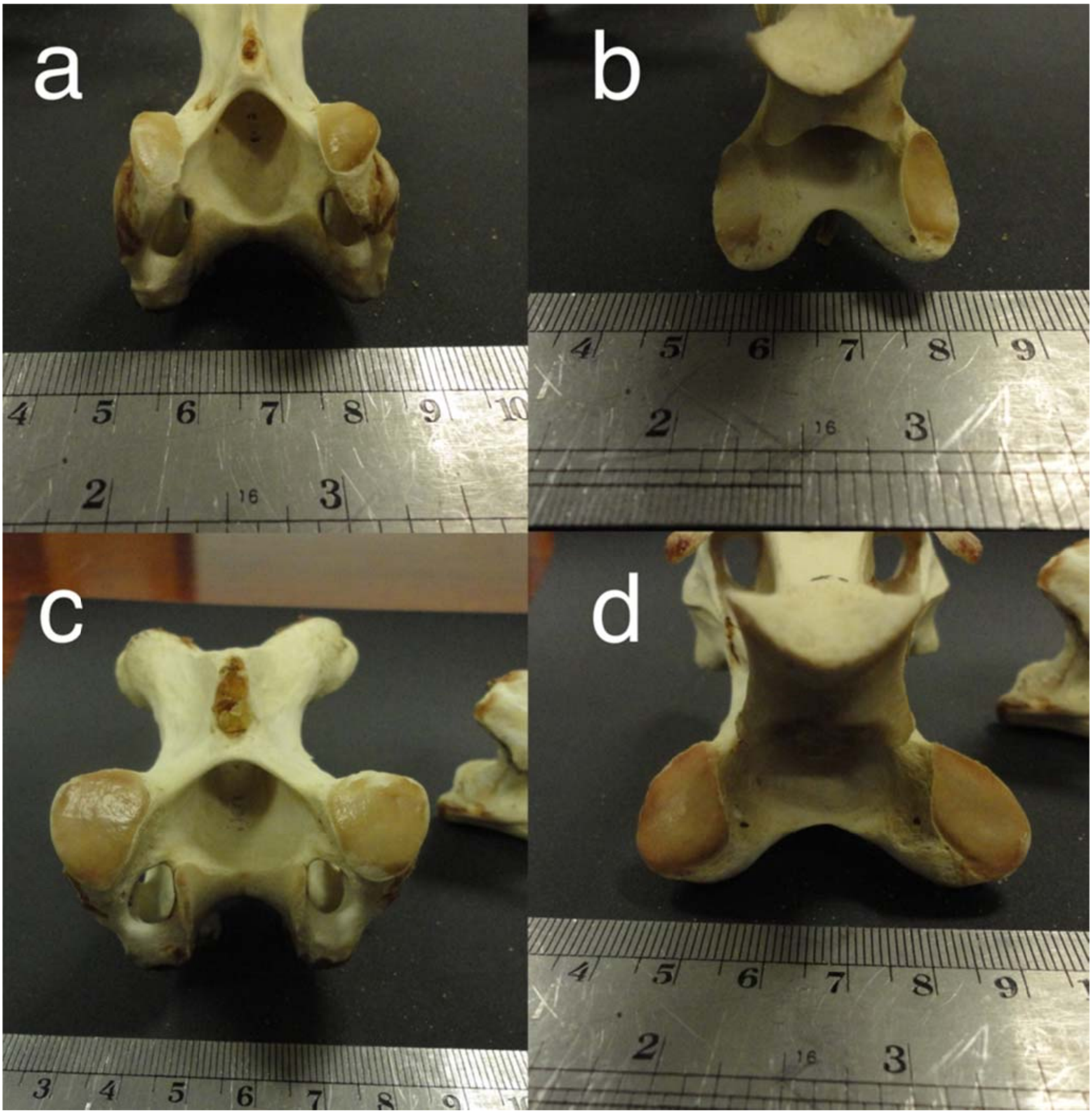

**Figure 16.**

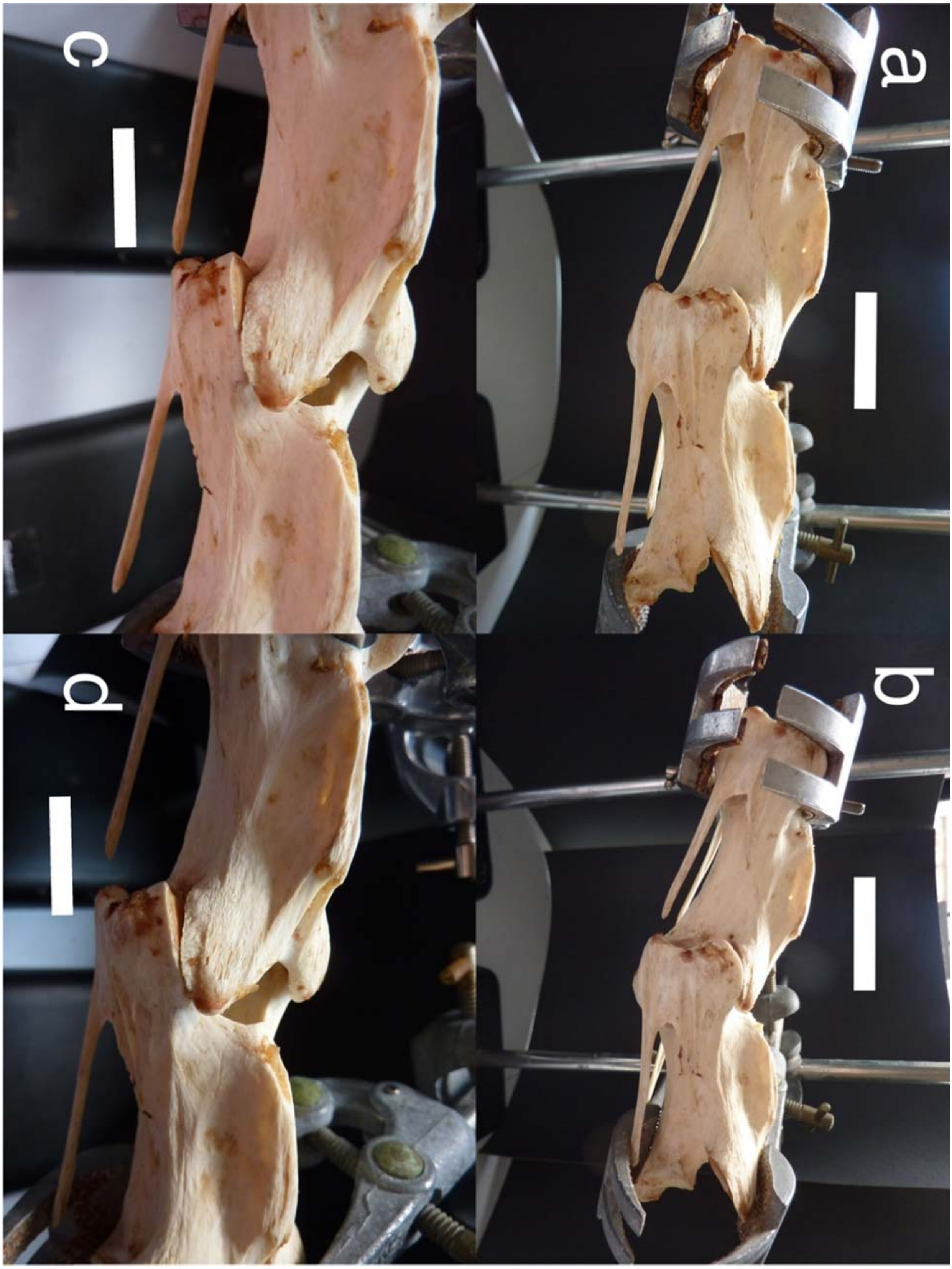

**Figure 17.**

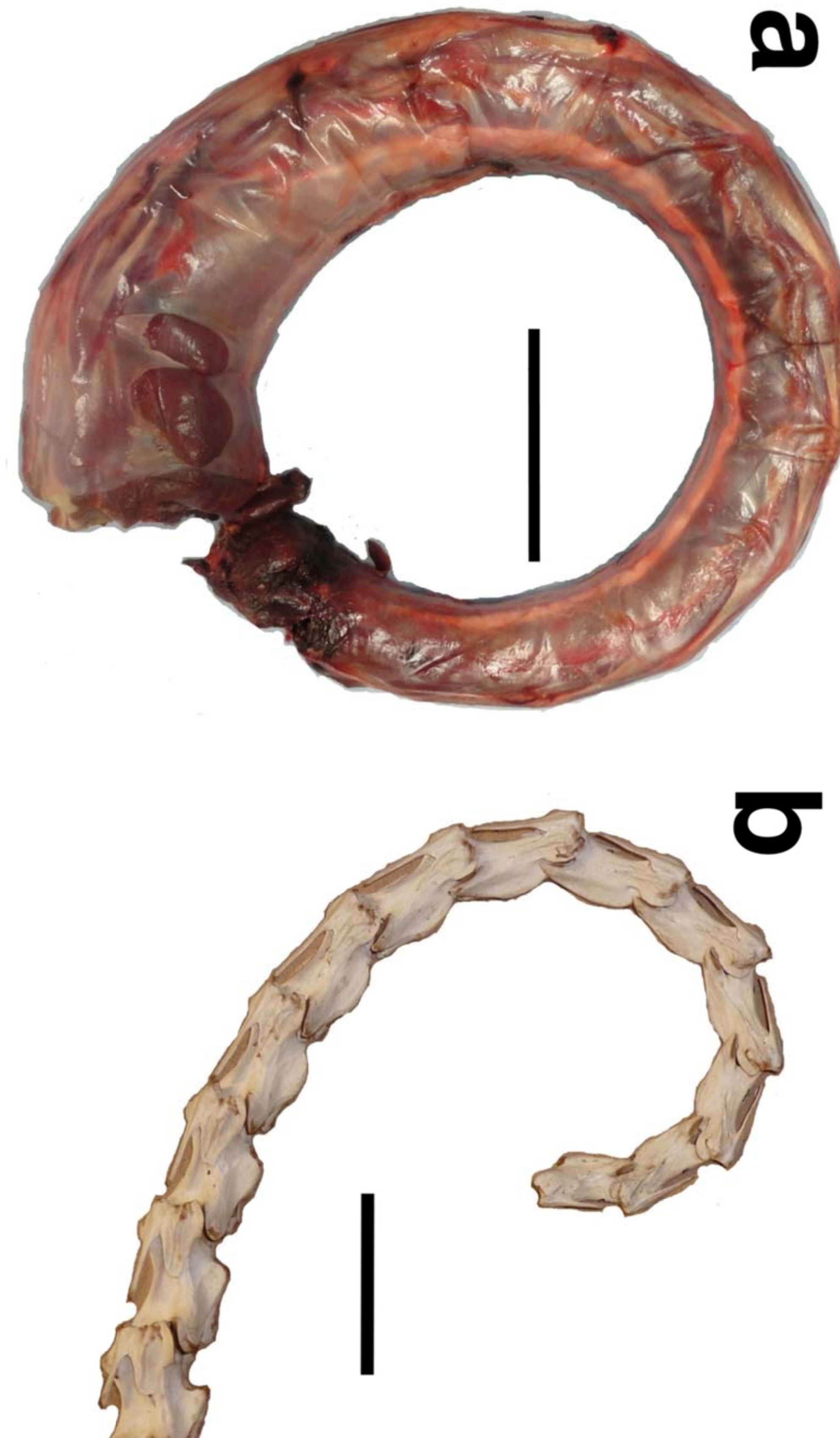

**Figure 18.**